\documentclass{optica-article}

\journal{opticajournal} 

\articletype{Research Article}

\usepackage{multirow}
\usepackage{array}
\usepackage{amsmath}
\usepackage{breqn}
\usepackage[title]{appendix}

\newcommand{\R}{}

\begin{document}

\title{Numerical simulation of coherent summation of laser beams in the presence of non-idealities in the dipole focusing system\footnote{Accepted for publication in Applied Optics.\\ DOI: https://doi.org/10.1364/AO.543161 \\ © 2024 Optica}}

\author{D. N. Bulanov, E. A. Khazanov, A. A. Shaykin, and A. V. Korzhimanov\authormark{*}}

\address{Federal Research Center A.V. Gaponov-Grekhov Institute of Applied Physics of the Russian Academy of Sciences, Nizhny Novgorod, Russian Federation\\
}

\email{\authormark{*}artem.korzhimanov@ipfran.ru} 


\begin{abstract*} 
A programming library was developed, based on Stratton-Chu diffraction integrals for calculating reflected optical fields. Dipole-type focusing schemes with tunable number of beams and mirror placements were studied, considering the influence of phase distortion and aberrations. The intensity above $3\times 10^{26}$ W/cm$^2$ was found theoretically attainable in a system of 12 beams of 50 PW each with about 90\% of that value realistically achievable.

\end{abstract*}

\section{Introduction}
The problem of obtaining high intensity radiation is of great interest for various fields of modern physics. One of the traditional approaches to solving this problem is the radiation amplification in a single channel. However, experimentally wise, further increase of the output power using this method is hindered by an effect of optical breakdown. One of the most promising and widespread alternatives is the coherent summation of several laser beams — this method is the key feature in projects \cite{ELI,XCELS} aimed at achieving peak powers of the order of several hundred PW and intensities of $\sim$$10^{25}$ W/cm$^2$.

To achieve the maximum possible value of the electromagnetic field in the system, a special approach to the focusing scheme design is also required. The dipole focusing scheme is theoretically shown \cite{bassett_limit_1986} to be optimal for monochromatic radiation of set power. That means the focused radiation should have the special structure of a time-reversed harmonic dipole radiation. This conclusion also holds for short laser pulses \cite{gonoskov_dipole_2012}. At low intensities the dipole wave structure could be obtained via profiling the intensity and polarization of a plane wave \cite{dorn_sharper_2003}. That is technically difficult to reproduce for the case of high powers (PW level), so it was proposed \cite{bulanov_multiple_2010} to create such a configuration from several pulses of limited aperture by placing them in a certain way using one or several (most commonly, off-axis parabolic) mirrors. For Gaussian beams, for example, such a scheme can consist of mirrors organized in two belts \cite{gonoskov_probing_2013} — that approximately simulates the power distribution of dipole wave, meaning the radiation from equatorial directions contributes to the focusing the most.

One of the projects, aimed at achieving high peak intensity values, is being developed at the Institute of Applied Physics of the Russian Academy of Sciences and is called XCELS (Exawatt Center for Extreme Light Studies) \cite{XCELS}. The design requires phase synchronization of 12 laser channels, each one providing up to 50 PW of power, bringing the total power to 600 PW and the intensity at the focus to values approaching the $\sim$$10^{26}$ W/cm$^2$. The field structure analysis at the proximity of the focus is one of the principal concerns for laser systems of this kind, because it defines the efficiency of coherent particle acceleration \cite{robinson_interaction_2018}. Additionally, electron-positron pair production, which is among the main research goals for XCELS, depends on the value of attainable intensity \cite{gonoskov_charged_2022}. Therefore, it is extremely important to investigate field distribution in focusing schemes and how various factors (for example, number of laser beams, position of optical elements and inaccuracies in their fabrication or alignment) impact the coherent beam summation efficiency. These studies are the main purposes of this article and they are performed via numerical simulation. Stratton---Chu integrals \cite{stratton_diffraction_1939} are used for calculations related to field reflection from curved mirrors. Section \ref{sec:focus} is devoted to their application and the description of the reflecting surface. Section \ref{sec:results} is divided into three parts and lists the results of numerical simulations, basing on which focusing schemes’ comparison is made, the coherent summation efficiency is investigated in various cases and the constraints are determined for each of them. The data are additionally compared with the results obtained earlier in a similar study for a system with quasi-parallel propagating Gaussian beams \cite{leshchenko_coherent_2015}.

\section{Focusing of a laser beam by an off-axis parabolic mirror}
\label{sec:focus}

Typically, working with high power lasers involves using off-axis parabolic (OAP) mirrors \cite{vais_direct_2018, vais_characterizing_2020}. \R{The key feature OAP mirrors are valued for is the ability to sustain very high optical quality regardless of the numerical aperture of the system}. Such a mirror is a part of a perfectly conducting paraboloid of revolution with the property of collecting all rays propagating parallel to the axis of the paraboloid towards the focus. Such a surface can be defined by the equation

\begin{equation}
z = \frac{x^2+y^2}{4F}-F
\label{Eq1}
\end{equation}
\R{In an arbitrary case, the expression has a different form from Eq.(\ref{Eq1}), but it can be reduced to an equivalent one by moving the coordinate origin and rotating the coordinate axes — after that, the same set of parameters is required to define the surface. The optical scheme corresponding to this configuration is shown in Fig.1.}

\begin{figure}[ht!]
\centering
\includegraphics[width=6cm]{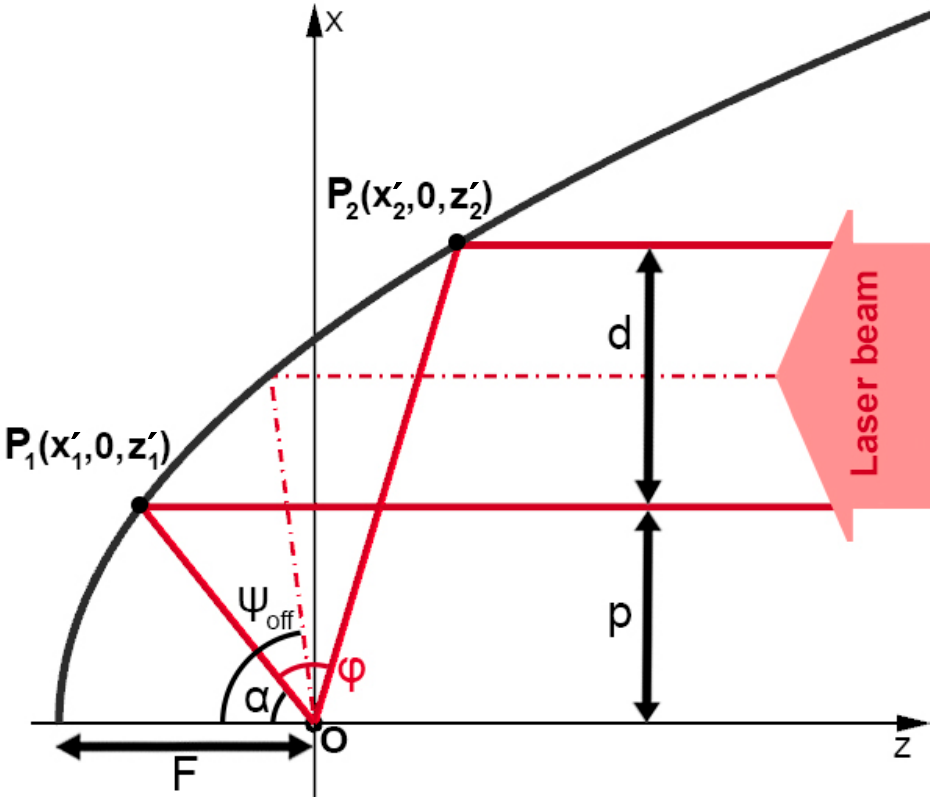}
\caption{Optical scheme of focusing laser beam by an OAP mirror. The following parameters are introduced: $\varphi$ is a convergence angle, $\alpha$ is a beam incidence angle (measured from a lower boarder of a beam), \textit{d} is a beam width, \textit{p} is an off-axis parameter, $\psi_{off}$ is an off-axis angle.}
\label{Fig1}
\end{figure}

In this case the focus is located in the point \textit{O}(0,0,0) at a distance \textit{F} from the vertex of the paraboloid. In reality, focusing of radiation strictly to a point is impossible and occurs in the area, which size depends on the beam parameters and conditions of incidence on the mirror. \R{For example, the Gaussian beam has the focusing spot size  given by}

\begin{equation}
d_{gauss} \approx \frac{4\lambda NM^2}{\pi}
\label{NewEq1}
\end{equation}
\R{where \textit{N} is numerical aperture of the system and $M^2$ is the beam quality parameter. For the case of monochromatic dipole wave, studied in \cite{gonoskov_dipole_2012}, the effective focal spot size area $\delta S_{md}$ was derived from} 

\begin{equation}
 \frac{1}{2}\frac{c}{4\pi}E_{max}^2 = \frac{P}{\delta S_{md}}
\label{NewEq2}
\end{equation}
\R{using average power \textit{P} and the maximum value of the electric field amplitude in a standing
\textit{e}-dipole wave $E_{max}$, and found to be approximately $0.12\lambda^2$, meaning the effective spot size is}

\begin{equation}
d_{md}\approx 2 \sqrt{\frac{0.12}{\pi}} \lambda \approx 0.39\lambda
\label{NewEq3}
\end{equation}

To calculate the fields after reflection and focusing we use Stratton---Chu integrals \cite{stratton_diffraction_1939}. If electric currents and charges are absent, $\varepsilon=1$ and $\mu=1$ these integrals are reduced to the forms

\begin{equation}
\vec{E}(x_1,y_1,z_1) = -\frac{1}{4\pi} \int_S {(ik[\vec{n} \times \vec{B}]\psi+[\vec{n} \times \vec{E}]\times\nabla\psi+(\vec{n} \cdot \vec{E})\nabla\psi)} \,dS
\label{Eq5}
\end{equation}

\begin{equation}
\vec{B}(x_1,y_1,z_1) = -\frac{1}{4\pi} \int_S {(-ik[\vec{n} \times \vec{E}]\psi+[\vec{n} \times \vec{B}]\times\nabla\psi+(\vec{n} \cdot \vec{B})\nabla\psi)} \,dS
\label{Eq6}
\end{equation}
Here $\vec{n}$ is a unit vector normal to the integration surface \textit{S}, $k=\frac{\omega}{c}$ is a laser wave number, $\psi=\frac{e^{ikr}}{r}$ is the Green’s function for the Helmholtz equation, \textit{r} is a distance between the observation point $M_1(x_1,y_1,z_1)$ and an arbitrary point $M(x,y,z)$ of the mirror, represented by surface \textit{S}

\begin{equation*}
r=\sqrt{(x-x_1)^2+(y-y_1)^2+(z-z_1)^2}
\end{equation*}

The mirrors are taken as ideally conducting surfaces. Denoting the fields in the incident and reflected wave as $\vec{E}_i$, $\vec{B}_i$ and $\vec{E}_r$, $\vec{B}_r$ respectively and considering the boundary conditions, we can get the following equations:

\begin{equation*}
\vec{E}(M)=\vec{E}_i+\vec{E}_r=2(\vec{n} \cdot \vec{E}_i)\cdot\vec{n}, \quad \vec{B}(M)=\vec{B}_i+\vec{B}_r=2[\vec{n} \times \vec{B}_i]\times\vec{n}
\end{equation*}
Substituting these expressions into Eq.(\ref{Eq5}), (\ref{Eq6}) allows to simplify them additionally:

\begin{equation}
\vec{E}(x_1,y_1,z_1) = \frac{1}{4\pi} \int_S {(2ik[\vec{n} \times \vec{B}_i]\psi+2(\vec{n} \cdot \vec{E}_i)\nabla\psi)} \,dS
\label{Eq7}
\end{equation}

\begin{equation}
\vec{B}(x_1,y_1,z_1) = \frac{1}{4\pi} \int_S {(2[\vec{n} \times \vec{B}_i]\times\nabla\psi)} \,dS
\label{Eq8}
\end{equation}

 Eq.(\ref{Eq7}), (\ref{Eq8}) contain information about the reflecting surface in the form of the integration area \textit{S} and can be conveniently transferred into programming code. For the description of the numerical simulation problem devoted to reflection of an arbitrary laser beam from a mirror of arbitrary shape see Supplementary.

\section{Results of numerical simulation}
\label{sec:results}
The numerical simulation is performed in the proximity of the system’s focus in two-dimensional areas of 24$\lambda$×24$\lambda$ in size in Section \ref{sec:results1} and three-dimensional areas of 8$\lambda$×8$\lambda$×8$\lambda$ in size in Sections \ref{sec:results2} and \ref{sec:results3}. These areas are discretized using 301×301 points in the first case and 81×81×81 points in the second case. The systems consist of an even number of laser beams (initially identical), focused by OAP mirrors in double-belt configuration. The characteristic size of the mirrors is $\sim$\textit{d} and $\gg$\textit{$\lambda$}. \R{The numerical aperture of the studied systems is in range between 1.04 and 1.81.} The beams are assumed to be linearly polarized with a square, super-Gaussian transverse profile \R{$E(x,y)\propto \exp\left\{-\left(x^{12}+y^{12}\right)/(d/2)^{12}\right\}$}, beam width \textit{d} is 20 cm. The power in a single beam is fixed to be 50 PW, and the wavelength $\lambda$ is 910 nm \cite{khazanov_exawatt_2023}. \R{We study the monochromatic case and expect the results to be applicable to the case of ultrashort pulses. This assumption is valid until pulse duration is at least 2--3 times longer than the light oscillation cycle \cite{christov_transmission_1985}.}

\R{To evaluate coherent summation efficiency in different cases we introduce the following coefficients:
the first one}

\begin{equation}
\eta_{\rm n}^{\rm des} = \frac{I_{\rm scheme_2}}{I_{\rm scheme_1}} 
\end{equation}
\R{is used to compare maximum achievable intensities for two focusing scheme designs with the same number of beams \textit{n}; the second one}

\begin{equation}
\eta_{\sigma} = \frac{I_{\rm N_2}}{I_{\rm N_1}} 
\end{equation}
\R{is used to compare maximum achievable intensities in one focusing scheme with the various number of beams ($N_2$ and $N_1$) with the jitter value of $\sigma$; and the third}

\begin{equation}
\eta_{\rm s} = \frac{I_{\rm dist}}{I_{\rm ideal}}
\label{NewEq6}
\end{equation}
\R{is defined as the ratio of the maximum field intensity in the focal region after coherent summation of "distorted" (influenced by aberrations or mirror misplacement) beams to the maximum intensity for the ideal (reference) case.}

In this paper, the following types of non-idealities in the focusing scheme are investigated:
 \begin{itemize}
     \item Phase difference (jitter) between incident beams;
     \item Inaccuracies in mirror alignment;
     \item Aberrations in the incident radiation;
 \end{itemize}
Similar studies \cite{leshchenko_coherent_2015, zhao_investigation_2015} have been carried out earlier for beams with Gaussian and super-Gaussian profiles of 4th and 10th orders respectively for focusing schemes different from those considered in the present work.

\subsection{Comparison of focusing schemes and the influence of jitter}
\label{sec:results1}
First of all, we evaluate the possibility of increasing the radiation intensity at the focus, using focusing schemes of various geometries and different number of beams, taking jitter into account. To do this we perform numerical simulation of electromagnetic fields in the focal area. It was done previously in \cite{gonoskov_probing_2013}, assuming Gaussian beams and fixed combined power of all beams. In our work, as mentioned above, the beams are super-Gaussian with square profile and we fix the power of a single beam. To study the influence of jitter, a random phase is added to each beam. The value of phase is generated from a normal distribution with a tunable width parameter, which is chosen the same for all beams within a single run of the program. The calculations are done 10 times for each width parameter, producing different sets of random phases.

\begin{figure}[ht!]
\centering
\includegraphics[width=12.5cm]{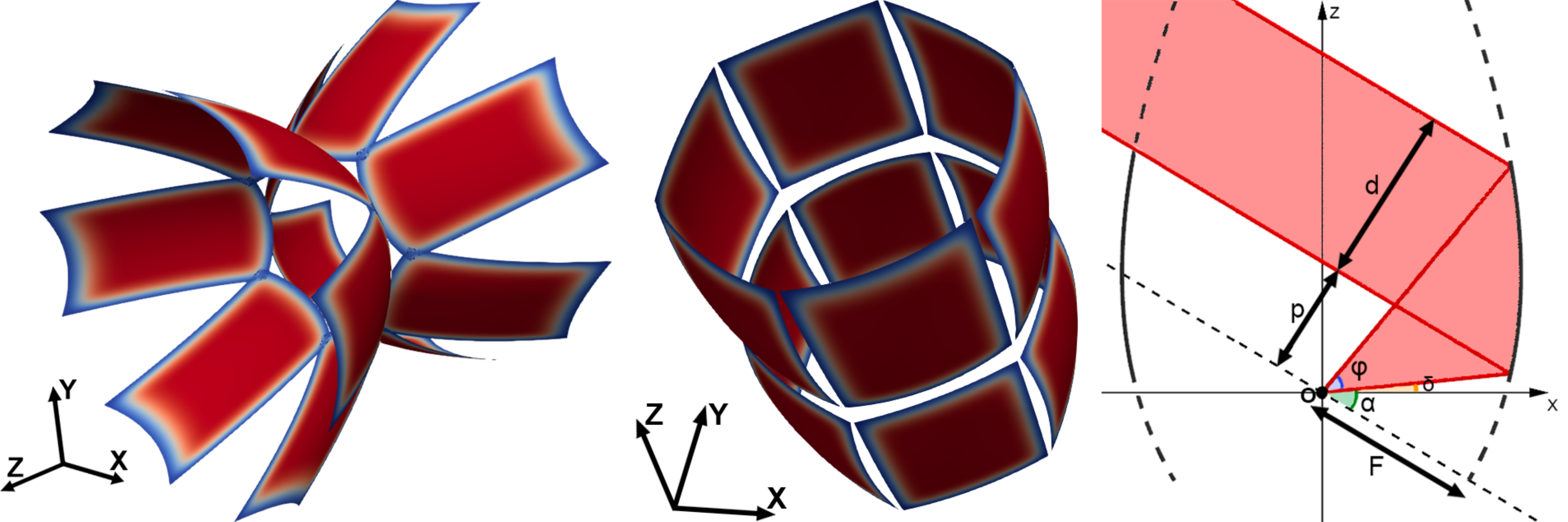}
\caption{Mirrors placement in focusing scheme 1 (left) and 2 (middle) for 12 beams (hereinafter, red areas mark the area of beam incidence, blue areas are the rest of the mirror). Focusing geometry of an OAP mirror in focusing scheme 2 (right)}
\label{Fig2}
\end{figure}

Two focusing scheme designs were investigated (Fig.\ref{Fig2}). The first one had been developed and studied earlier for Gaussian beams and fixed total power \cite{gonoskov_probing_2013}. It assumes that the mirrors are grouped together, with the lowest point touching the equator line and the lower corners touching each other. The beams propagate along the polar axis and the shape of mirrors (focal length of the paraboloid) is chosen accordingly to a number of beams. However, due to chosen beam propagation direction, the beam aperture achieved is not optimal.

The second scheme was designed taking into account certain experimental purposes. The configuration supposes that beams propagate in such a way, that their lower boundary touches the upper boundary of the mirrors on the opposite side — in this case the mirrors end up arranged quasi-vertically. This design has been developed for the XCELS project. Additionally, the presence of technological gaps between the mirrors to access the focus for the purpose of diagnostics and alignment is provided — to take this into account, an additional parameter $\delta$ is used during mirror model definition.

The intensity \textit{I} in the focus of a dipole focusing scheme is defined as

\begin{equation}
I =
 \frac{cE^2}{8\pi}
 \label{eq12}
\end{equation}
where \textit{c} is a speed of light, \textit{E} is an amplitude of the electric field. It has been shown earlier \cite{bassett_limit_1986, dorn_sharper_2003} that for a given total power of radiation \textit{P} the theoretical maximum of intensity is achieved by dipole wave and it equals to $I_{\rm dipole} = {8\pi P}/{3\lambda^2}$ where $\lambda$ is a wavelength.

\begin{figure}[ht!]
\centering
\includegraphics[width=6cm]{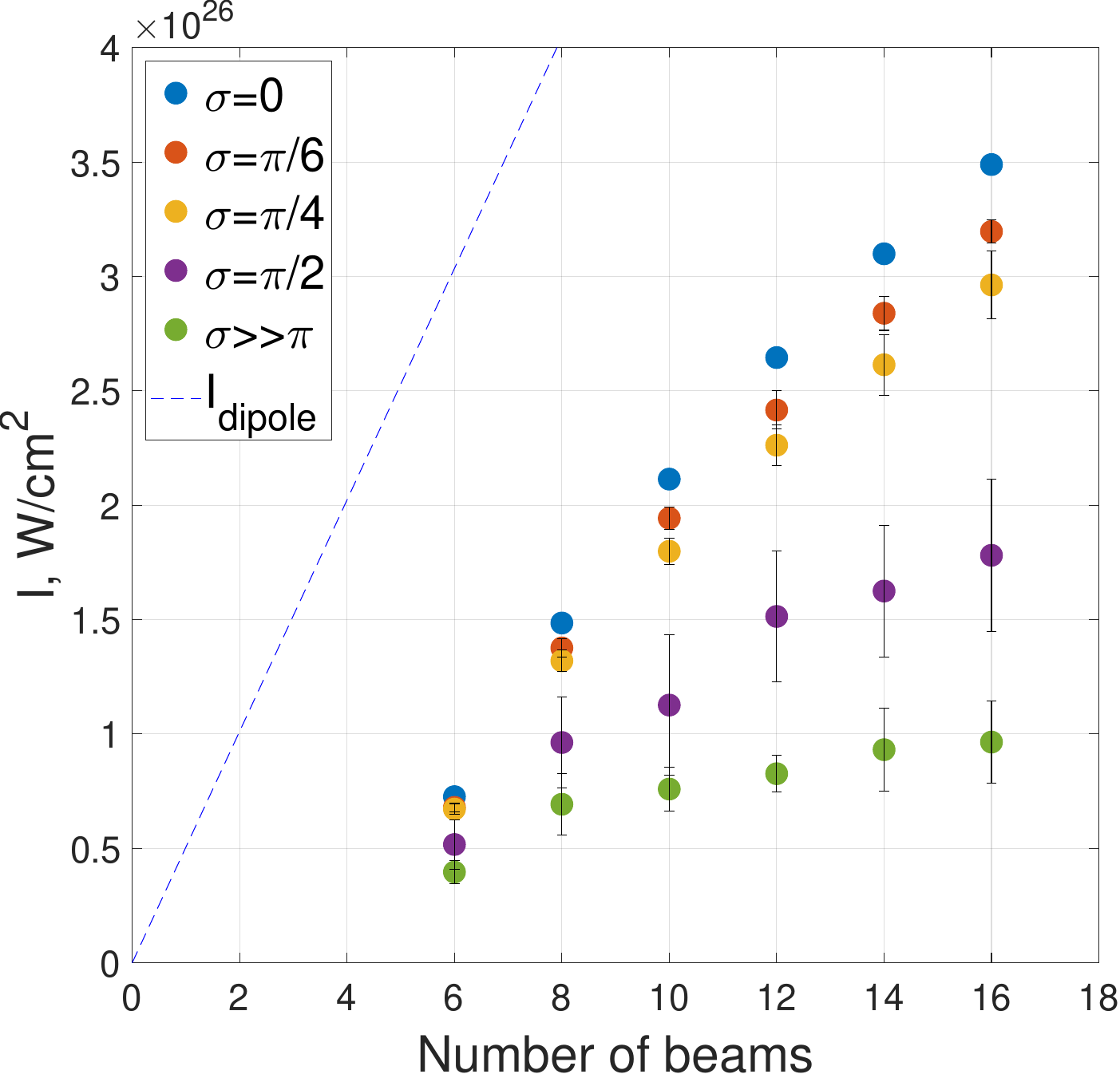}
\includegraphics[width=6cm]{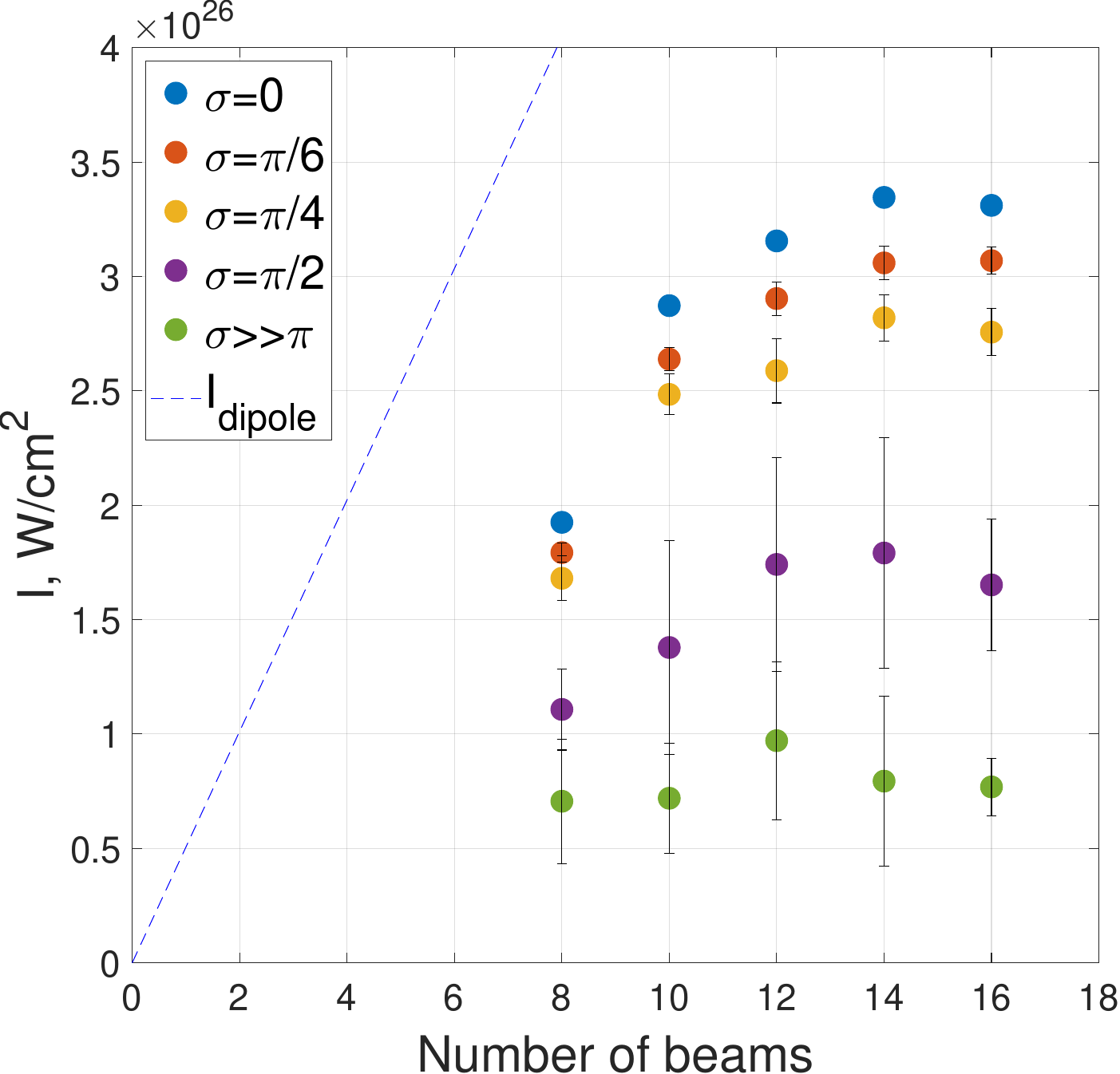}
\caption{Dependencies of the maximum intensity (averaged over 10 runs) achieved in the focal region (not necessarily in the geometrical center of the system) on the number of beams for ideal case and different jitter values for focusing scheme 1 (left) and 2 (right)}
\label{Fig4}
\end{figure}

To evaluate the efficiency of represented focusing schemes in obtaining high intensity values and the influence of jitter the calculations are done for different number of beams. A random phase was generated from a normal distribution with a set standard deviation $\sigma$. \R{The results of corresponding calculations are presented in Fig.\ref{Fig4}.}

In the case of the first focusing scheme the dependence of the maximum intensity on the number of beams is quasilinear, whereas it is quadratic for a coherent summation of plane waves. Due to the shape of the beam profile, the convergence angle depends on the number of beams — the more beams are used, the smaller the equatorial convergence angle and the smaller the polar convergence angle. Overall, the presence of jitter leads to the worsening of coherent summation efficiency: at the values of around $\frac{\pi}{2}$ switching \R{from 8 to 14 beams results in intensity increase from 0.96×$10^{26}$ to 1.63×$10^{26}$ W/cm$^2$ ($\eta_{\pi / 2}\approx 1.7$), while in ideally synchronized case the values are 1.49×$10^{26}$ and 3.09×$10^{26}$ W/cm$^2$ for 8 and 14 beams respectively ($\eta_{0}\approx 2.07$)}.

Considering the second focusing scheme, it is proven to be \R{more preferable for a certain range of beams number} as it allows to achieve higher peak intensity values \R{($\eta_{8}^{\rm des}\approx 2.64$, $\eta_{10}^{\rm des}\approx 1.93$, $\eta_{12}^{\rm des}\approx 1.49$, $\eta_{14}^{\rm des}\approx 1.08$)}. The dependence of the maximum intensity on the number of beams is also not quadratic — it tends to have an optimum, as polar angle of focusing decreases with increase of beam number, so the focal spot becomes larger. The maximum intensity is 3.15×$10^{26}$ W/cm$^2$ for 12 beams, which is a design configuration of XCELS. That is 6\% less, than the optimum value, achieved at 14 beams (3.35×$10^{26}$ W/cm$^2$), and is about 50\% of theoretical value for a dipole scheme. The presence of jitter also makes beam summation less effective — the phase of around $\frac{\pi}{2}$ leads to \R{intensity increase from 1.11×$10^{26}$ to 1.79×$10^{26}$ W/cm$^2$ ($\eta_{\pi / 2}\approx 1.61$), switching from 8 to 14 beams, whereas for ideal synchronization the values are from 1.92×$10^{26}$ to 3.35×$10^{26}$ W/cm$^2$ respectively ($\eta_{0}\approx 1.74$).}

\R{The detailed intensity distribution calculated in focal area for both focusing schemes with 12 beams is given in Supplementary. Fig.\ref{NewFig1} is devoted only to the second scheme in ideal case and under the influence of various inaccuracies, discussed further in more detail.}

\begin{figure}[ht!]
\centering
\includegraphics[width=13cm]{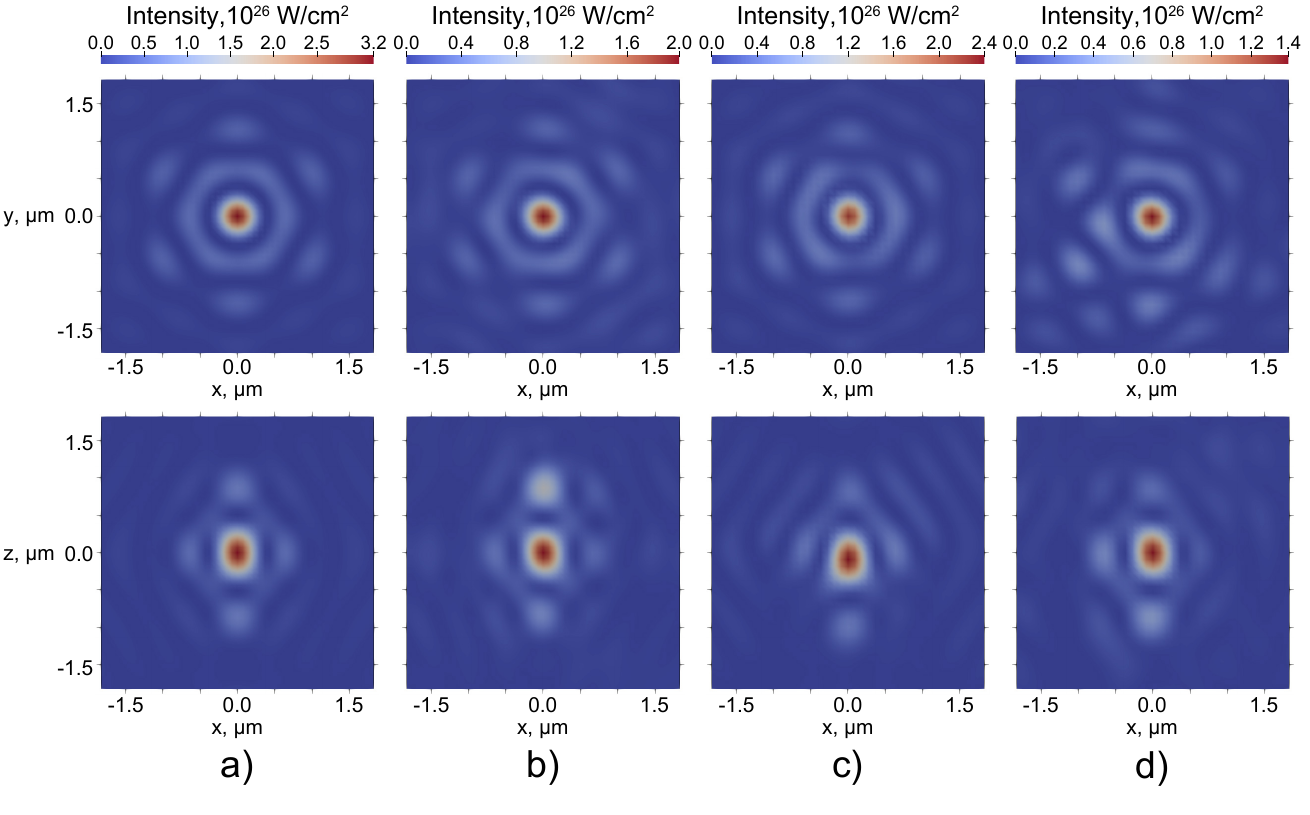}
\caption{Intensity distribution in focal area of a system with 12 beams for equatorial (top) and
meridional (bottom) planes in case of a) ideal beam phasing and mirror alignment; b) mirror shifts ($\sigma$=0.5 $\mu$m); c) mirror rotations ($\sigma$=5×$10^{-7}$ rad.); d) aberrations ($\sigma_{beam}\approx$ 0.83 rad.) for the focusing scheme 2}
\label{NewFig1}
\end{figure}
\R{Using calculated intensity 3.15×$10^{26}$ W/cm$^2$ and Eq.(\ref{NewEq2}) for the case of 12 beams, we can estimate the size of focal spot to be approximately $0.54\lambda$. In comparison, the result for super-Gaussian beam of 12th power and equivalent mirror ($N \approx 1.29$, $M^2 \approx 1.74$), using Eq.(\ref{NewEq1}), is $2.85\lambda$. And even in the case of Gaussian pulse ($M^2 = 1$) it is $1.64\lambda$. We see that the multi-mirror configuration is much closer to ideal monochromatic dipole wave, for which the effective focal spot is  $0.39\lambda$, given by Eq.(\ref{NewEq3})}

We can compare the results presented in this section to those obtained in \cite{leshchenko_coherent_2015}, where the study was carried out for the quasi-parallel propagation of focused beams. For such a configuration, it was concluded that the intensity decreases by 10\% when the phase mismatch is around 0.5 rad. for 8 beams and 0.4 rad. for 10 or more beams. \R{In contrast, our study describes the situation, when focused beams are counter propagating.} For both focusing schemes we estimated that 90\% of maximum intensity value is achievable at a mismatch of around 0.7 rad. for 8 beams and 0.5 rad. for 10 or more beams. In other words, \R{we assume that} in systems with counter propagating beams, the presence of jitter has a weaker effect on the coherent summation than in systems with quasi-parallel propagation. \R{There can be the following explanation} — when two plane waves move towards each other, \R{the relative phase delay does not change the maximum intensity — instead it is just reached in a different point; for the case of two plane waves propagating in the same direction the phase delay $\Delta\varphi$ decreases the maximum intensity by $\cos^2\Delta\varphi$}.

Further investigations are carried out for the second focusing scheme and for 12 beams \R{(numerical aperture is $\approx1.29$)}. This case, presented in Fig.\ref{Fig2} (middle), is considered a reference one, meaning all mirror movements are done relatively to their location in this scheme, and the obtained value of the peak intensity (3.15×$10^{26}$ W/cm$^2$) is used to evaluate the efficiency of coherent summation \R{(note Eq.(\ref{NewEq6}))} and referred to as $I_{\rm ideal}$.

\subsection{Influence of mirror alignment inaccuracies}
\label{sec:results2}

\begin{figure}[ht!]
\centering
\includegraphics[width=13cm]{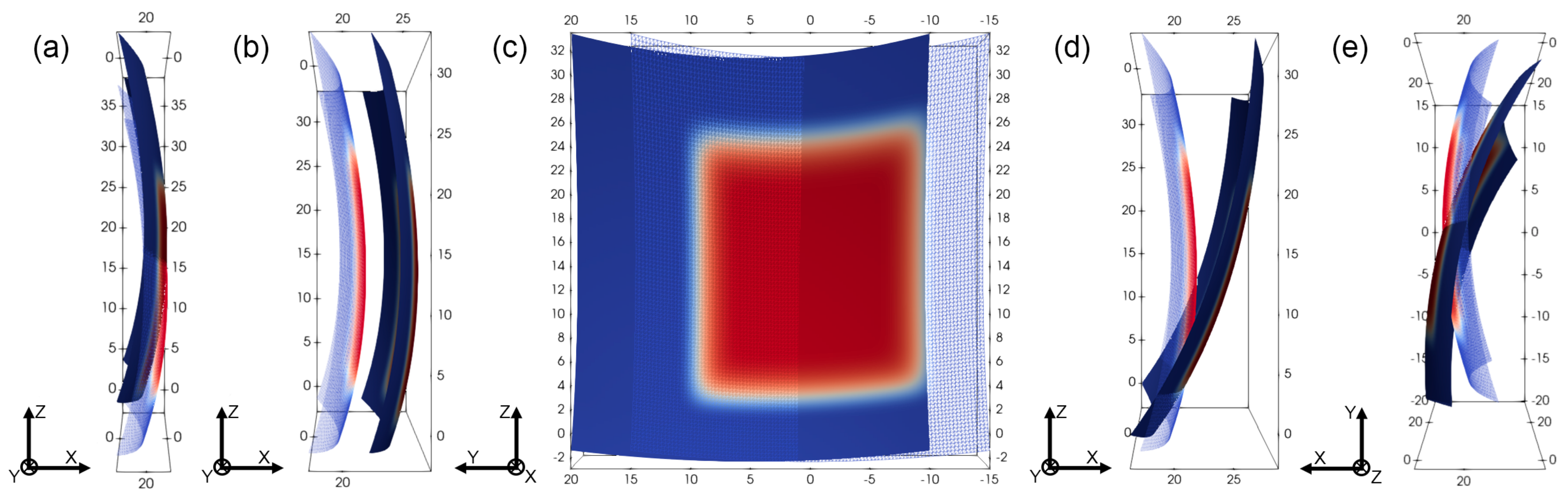}
\caption{Mirror shifts in vertical (a) and horizontal (b,c) directions. Mirror rotations around \textit{y} (d, type 1) and \textit{z} (e, type 2) axes. The reference mirror position in focusing scheme is shown as semi-transparent.}
\label{Fig5}
\end{figure}

The influence of inaccuracy in mirror alignment on coherent summation of laser beams is investigated by modeling various kinds of mirror deviations from the reference position while maintaining the propagation directions of incident beams. Two types of mirror position changes are studied — shifts and rotations (Fig.\ref{Fig5}). There are three kinds of shifts simulated — one vertical and two horizontal (along the radial direction and along the tangent line to the circle, which the mirrors in the belt form). Additional calculations are performed for rotations around \textit{y} and \textit{z} axes — they are labeled as type 1 and 2 accordingly.

To evaluate the coherent summation efficiency in these cases the calculations are performed 10 times for random sets of shifts and rotation angles, generated from a normal distribution with a tunable standard deviation $\sigma$. Each mirror is moved independently in a single run.

\begin{figure}[ht!]
\centering
\includegraphics[width=6cm]{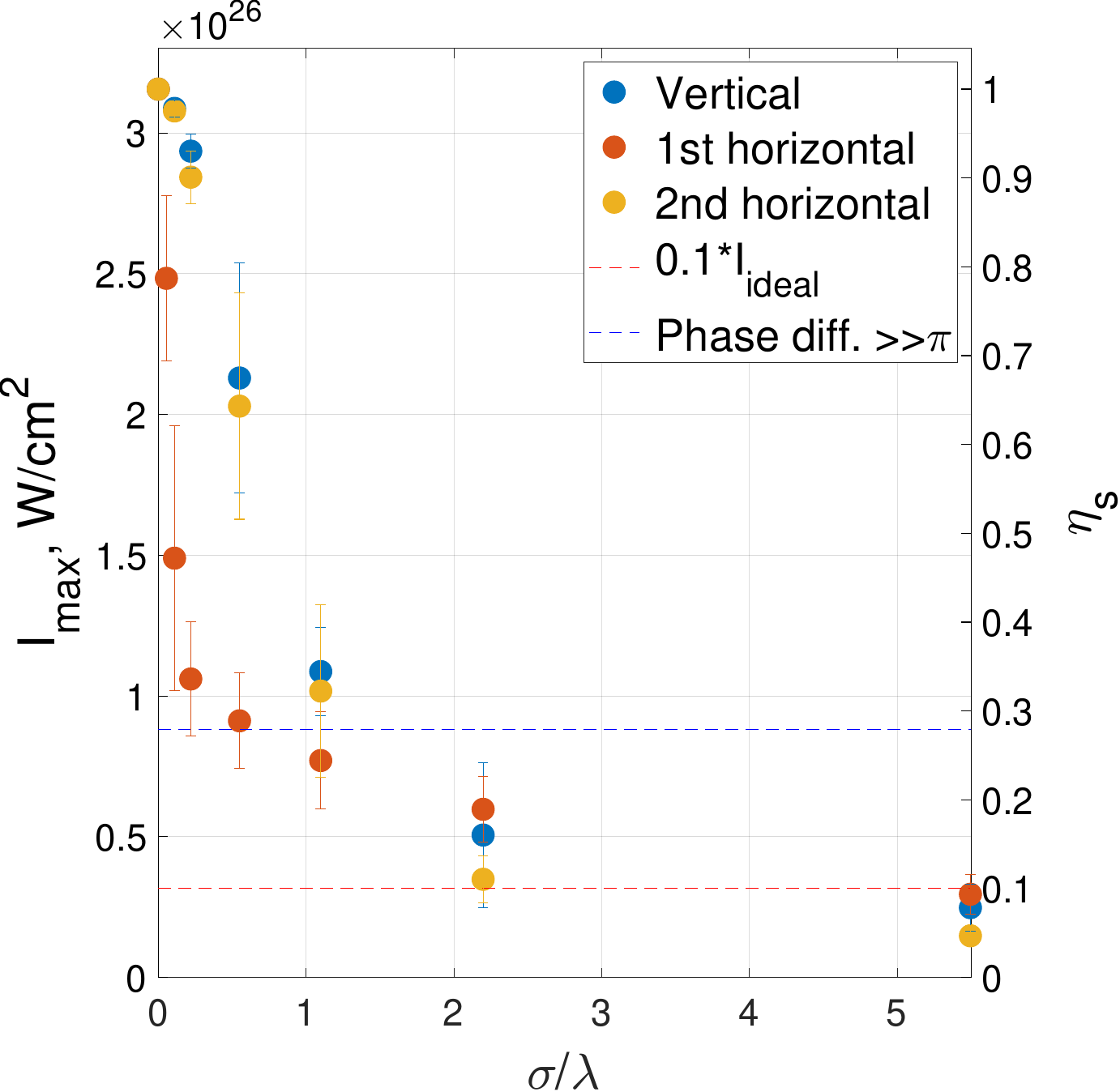}
\includegraphics[width=6cm]{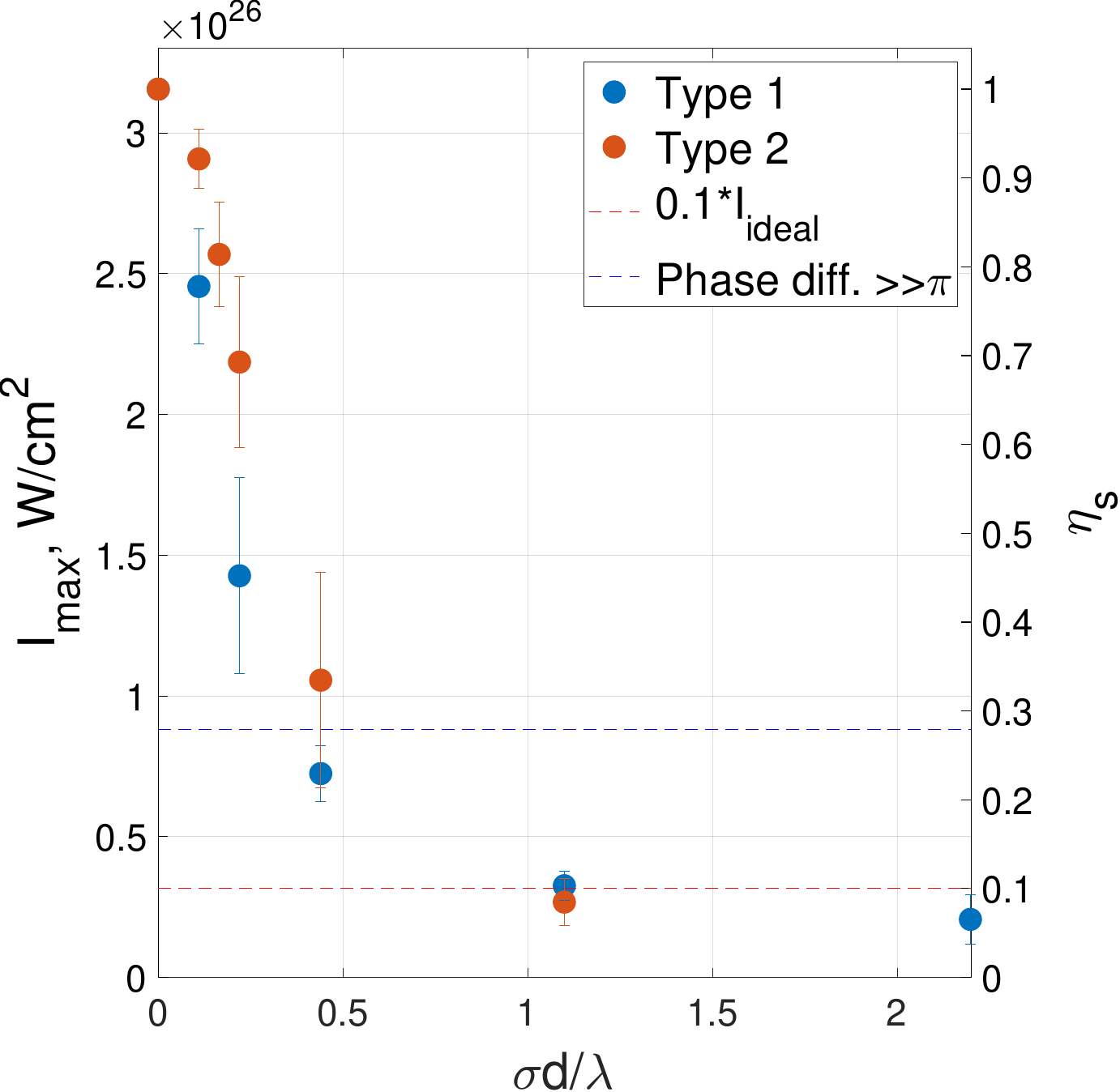}
\caption{Dependence of the maximum intensity (averaged over 10 runs) achieved in the focal region and parameter $\eta_{\rm s}$ on the $\sigma$-related parameter for shifts (left) and rotations (right)}
\label{Fig7}
\end{figure}

It should be noted, that in the case of shifts the maximum intensity and parameter $\eta_{\rm s}$ are taken as functions of $\frac{\sigma}{\lambda}$, whereas in the case of rotations they are presented as functions of parameter $\frac{\sigma d}{\lambda}$. That is due to the fact, that, when the mirror is simply shifted, the point of focus is moved by the distance of $\sigma$ (measured in µm), while when the mirror is rotated, the point of focus is moved by the distance of $\sigma d$ (here $\sigma$ is measured in rad.), meaning the dimensionless parameter is $\frac{\sigma d}{\lambda}$ instead of $\frac{\sigma}{\lambda}$.

The results are shown in Fig.\ref{Fig7}. It can be noted that the dependencies are characteristically similar for both types of misalignments. For vertical and tangent horizontal shifts intensity values coincide within the margin of error, whereas for radial horizontal shifts the decline is much sharper at the initial stage  — that can be explained by the fact that when the mirror is moved from/to the center of the system, the focus is shifted in both horizontal and vertical directions simultaneously, while the other movements change the position only along one of the directions. There is also a similar difference for the type 1 and 2 rotations. Since the bottom side of the beam on the mirror was chosen as the rotation axis in the first case, than, on average, the deviation from the reference position for the points of the mirror in the beam region will be more significant compared to the second case, when the axis passes through the middle of the beam.

To summarize, a noticeable drop in intensity occurs when the focus point is moved by half a wavelength: for vertical and tangent horizontal shifts the intensity decreases by 30\%, for the radial shift and in the case of rotations — by 70\%. When the mirror position is changed by an order of wavelength, the maximum achievable intensity is $\sim$10 times smaller compared to the reference case. To maintain a level of 90\% of the reference intensity value, the mirror should not be displaced by more than 0.2$\lambda$ for shifts and 0.1$\lambda$ for rotations \R{(meaning the misplacement of the edge of the mirror after the rotation)}.

\subsection{Influence of wavefront aberrations}
\label{sec:results3}

\begin{figure}[ht!]
\centering
\includegraphics[width=7.5cm]{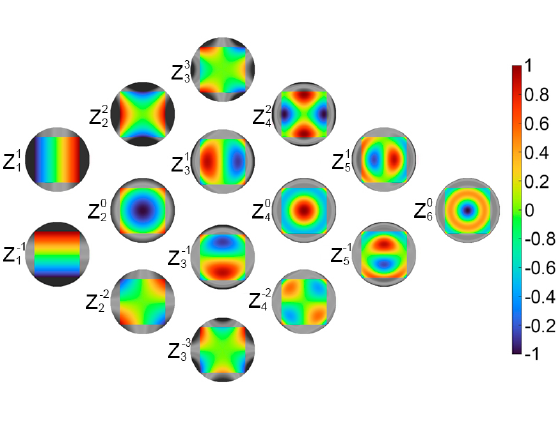}
\includegraphics[width=5.5cm]{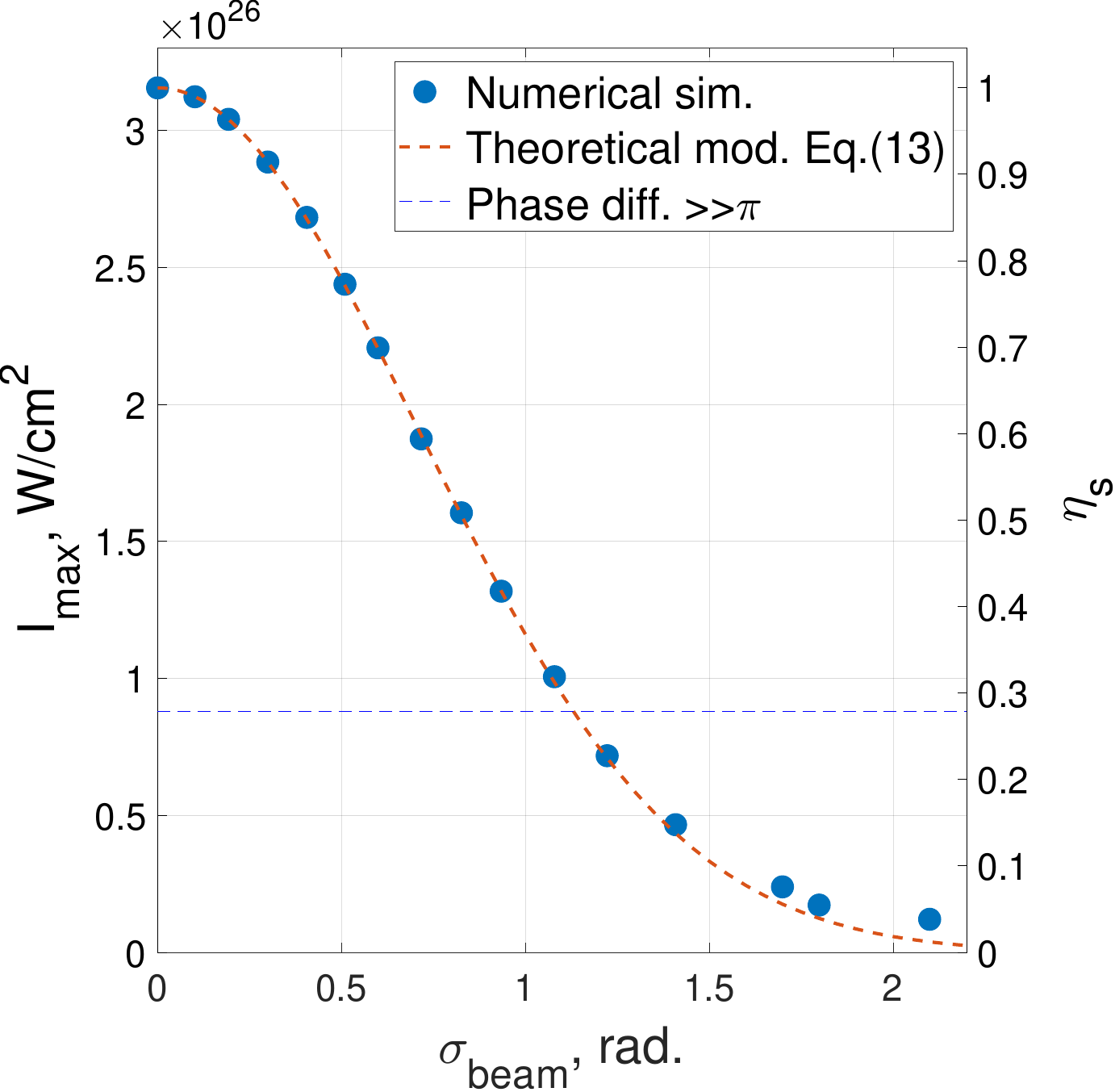}
\caption{Phase distortions introduced by Zernike polynomials used for numerical calculations (the areas of simulation are plotted in color) (left). Dependence of the maximum intensity (averaged over 10 runs) achieved in the focal region and parameter $\eta_{\rm s}$ on RMS aberration value $\sigma_{beam}$ (averaged over 10 runs and all beams) for aberrations given by 26 Zernike polynomials with \textit{n}=1..6 (right)}
\label{Fig8}
\end{figure}

When simulating aberrations in the incident radiation, both cumulative and isolated influence of the selected set of distortion harmonics on the efficiency of coherent addition is investigated. To describe the wavefront aberrations, the Zernike model \cite{born_principles_1999} is used — a multiplier containing one or more Zernike polynomials $Z_j(x_1,x_2 )=Z_n^m(\rho,\phi)$ with tunable amplitude $A_j$ is introduced into the beam. Here, $x_1$ and $x_2$ are coordinates responsible for the spatial profile of a beam (they should be normalized by half of the beam width). It is worth to mention that Zernike polynomials\R{, being pretty common and recognizable model in optics,} are used mainly for circular profiles (like Gaussian), while we study a beam profile that is close to a square shape. \R{To somehow compensate for that fact,} the coordinates were renormalized by half the length of the square diagonal, because we need the aberrations to cover the entire beam profile (normalization by the half-width of the beam results in the "corners" of the square profile remaining unchanged). However, in this case areas at the edges of the unit circle, which the Zernike polynomials are defined on, are, in fact, cut off (see example in Fig.\ref{Fig8} (left)). \R{After these manipulations, the majority of polynomials used in simulations remain linearly independent, therefore we suppose that such description is viable enough.}

To evaluate the coherent summation efficiency the calculations are performed 10 times for random sets of amplitude values $A_j$, generated from a normal distribution with a tunable standard deviation $\sigma$. These amplitude values are responsible for RMS aberration value of a single beam $\sigma_{\rm beam}$. One or more factors $A_j$ are generated for each beam independently for a single run.

The efficiency of coherent summation of 12 beams with a phase multiplier containing the first 27 Zernike polynomials is investigated, excluding $Z_0^0$ because it is responsible for jitter and has been already discussed. Theoretically, we expect the following: the aperture of each beam can be divided into a large number of sub-apertures with a constant phase (within the sub-aperture), then the field at the focus is the sum of fields from all sub-apertures, and each summand has its own phase. With a large enough number of Zernike polynomials and beams, it is reasonable to assume that the phase of the field in the focus from a single sub-aperture is a random variable with a normal distribution law, zero mean and dispersion value $\sigma_{\rm beam}^2$, equal to the square of the RMS aberration value at the beam aperture. Then the parameter $\eta_{\rm s}$, averaged over the ensemble of simulations, should follow

\begin{equation}
\eta_{\rm s} = e^{-\sigma_{\rm beam}^2}
\end{equation}
The obtained data, shown in Fig.\ref{Fig8} (right), matches almost perfectly to a described theoretical model. It can be concluded that to achieve 90\% efficiency the value of RMS aberrations should not exceed 0.3 radians, which is potentially achievable under realistic conditions using adaptive optics \cite{soloviev_improving_2022}. The intensity achieved in that scenario ($\sim$2.8×$10^{26}$ W/cm$^2$) is still higher than the value that is possible to reach for the focusing scheme with quasi-parallel beam propagation (Fig.\ref{Fig4} (left)) even in ideal situation ($\sim$2.65×$10^{26}$ W/cm$^2$).

\begin{figure}[ht!]
\centering
\includegraphics[width=7.5cm]{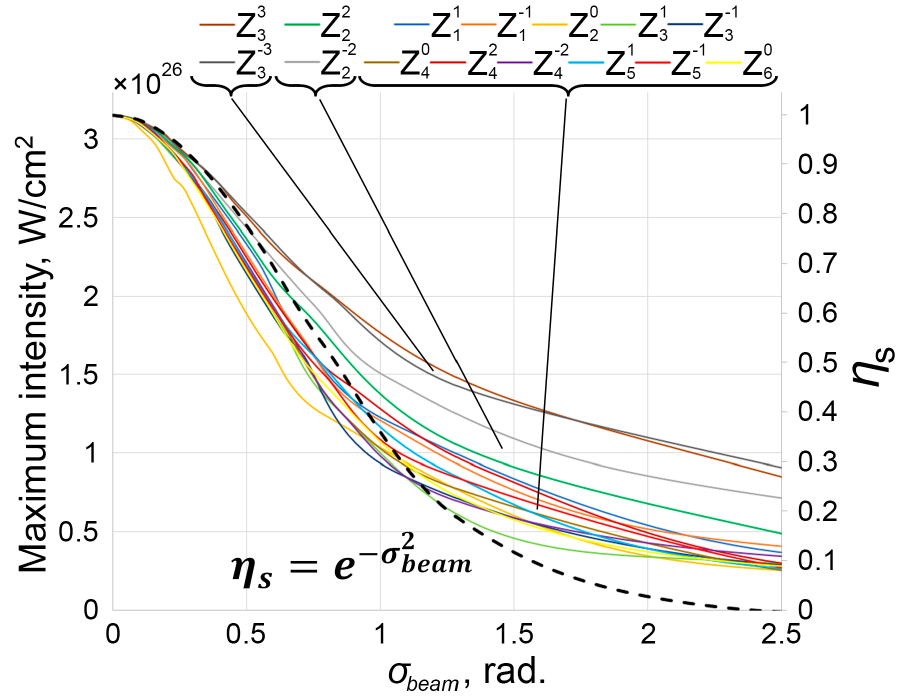}
\caption{Dependence of the maximum intensity (averaged over 10 runs) achieved in the focal region and parameter $\eta_{\rm s}$ on RMS aberration value $\sigma_{beam}$ (averaged over 10 runs and all beams) for aberrations defined by individual Zernike polynomials}
\label{Fig10}
\end{figure}

 \R{It is important to mention that the cumulative effect of a set of aberrations depends on their distribution (meaning the result above is applicable to the situation, when the aberrations are uniformly distributed). That is why the isolated effect of different types of aberrations should be studied additionally.} In this case, a phase multiplier contains single Zernike polynomial $Z_n^m(\rho,\phi)$, every one of which corresponds to a certain aberration harmonic. All types of aberrations investigated are presented in Fig.\ref{Fig8} (left). The results of numerical simulation are plotted in Fig.\ref{Fig10}.

Qualitatively, the dependence is the same as ones obtained in Section \ref{sec:results2}. It can be concluded that in order to achieve 90\% efficiency of coherent summation the value of RMS aberration should not exceed 0.2--0.4 radians depending on the type of aberration, \R{therefore we suppose they influence the coherent summation and focusing in general differently. Similar effects were described previously in \cite{leshchenko_coherent_2015} and \cite{rondepierre_propagation_2023} for other experimental setups.}

We can compare the results of this section to those obtained in \cite{leshchenko_coherent_2015}, where the study was performed using a similar approach to describe aberrations. It is worth to mention, that the conclusions in \cite{leshchenko_coherent_2015} were found to be fair independently of the number of beams in the system. First of all, primary coma aberrations $Z_3^{\pm 1}$ were listed as the least impactful in terms of coherent summation efficiency. That is not what happens in our configuration, supposedly because primary coma aberrations change the phase on the lower part of the beam (which gives the main contribution to the field at the focus of the system) prominently. Secondly, trefoil-type aberrations ($Z_3^{\pm 3}$), primary ($Z_2^{\pm 2}$) and secondary astigmatism ($Z_4^{\pm 2}$) were mentioned to give the next highest contributions in \cite{leshchenko_coherent_2015}. In this respect, their effect is similar in our results, therefore we assume that the polynomials $Z_n^m(\rho,\phi)$ with a larger absolute value of the upper index \textit{m} have a weaker effect on the coherent summation efficiency. Finally, the constraint on achieving 90\% efficiency for the value of RMS aberration was shown to be 0.2--0.6 radians. Based on our calculations, the estimated range is 0.2--0.4 radians, i.e. the system is more sensitive to the presence of aberrations.

\section{Conclusion}

In our work we studied the process of coherent summation of laser beams in a quasi-dipole configurations and obtained results based on the numerical simulation data. We evaluated the possibility to increase the intensity in the focal area, using different number of beams and focusing schemes. The key result is that the intensity of above 3×$10^{26}$ W/cm$^2$  is attainable in a system of 12 beams of 50 PW each (focusing scheme 2) — this configuration corresponds to the design characteristics of XCELS \cite{khazanov_exawatt_2023}. The following constraints were determined to be imposed in order to achieve at least 90\% of the intensity value of the ideal configuration:
 \begin{itemize}
     \item the phase difference between beams shall not exceed 0.7 radians for 8 beams and 0.5 radians for 10 or more beams;
     \item the mirror should not be displaced by more than 0.2$\lambda$ for shifts and 0.1$\lambda$ for rotations;
     \item the RMS aberration value should not exceed 0.3 radians \R{in a single beam}.
 \end{itemize}
\R{Despite being challenging,} all the conditions above \R{seem to be} achievable at the current level of technological development (cf. \cite{fu2024}). The programming library, developed for research purposes, could be used to solve the wide range of problems related to calculations of propagation and reflection of electromagnetic waves, which are beyond the scope of this work. \R{Possible future studies involve numerical simulation of focusing the single-cycle pulses and comparing the calculation results to the results of the experiments with two beams.}

\begin{backmatter}

\bmsection{Acknowledgments}
This research was supported by Center of Excellence ”Center of Photonics” funded by The Ministry of Science and Higher Education of the Russian Federation, contract 075-15-2020-906. The authors thank Maria Vernikovskaya and Aleksey Bulatov for their help with the developing of the numerical code. The simulations were performed on resources provided by the Joint Supercomputer Center of the Russian Academy of Sciences.

\end{backmatter}

\bibliography{refs}

\begin{thebibliography}{10}
\newcommand{\enquote}[1]{``#1''}

\bibitem{ELI}
\enquote{{ELI} — {Extreme} {Light} {Infrastructure}, {White} book,} \url{https://eli-laser.eu/media/1019/eli-whitebook.pdf}.

\bibitem{XCELS}
\enquote{{Exawatt} {Center} for {Extreme} {Light} {Studies} ({XCELS}), {Project} {Summary},} \url{https://xcels.ipfran.ru/img/site-XCELS.pdf}.

\bibitem{bassett_limit_1986}
I.~Bassett, \enquote{Limit to {Concentration} by {Focusing},} {\protect\JournalTitle{Optica Acta: International Journal of Optics}} \textbf{33}, 279--286 (1986).

\bibitem{gonoskov_dipole_2012}
I.~Gonoskov, A.~Aiello, S.~Heugel, and G.~Leuchs, \enquote{Dipole pulse theory: {Maximizing} the field amplitude from 4$\pi$ focused laser pulses,} {\protect\JournalTitle{Phys. Rev. A}} \textbf{86}, 053836 (2012).

\bibitem{dorn_sharper_2003}
R.~Dorn, S.~Quabis, and G.~Leuchs, \enquote{Sharper {Focus} for a {Radially} {Polarized} {Light} {Beam},} {\protect\JournalTitle{Phys. Rev. Lett.}} \textbf{91}, 233901 (2003).

\bibitem{bulanov_multiple_2010}
S.~S. Bulanov, V.~D. Mur, N.~B. Narozhny, \emph{et~al.}, \enquote{Multiple {Colliding} {Electromagnetic} {Pulses}: {A} {Way} to {Lower} the {Threshold} of $e^+ e^-$ {Pair} {Production} from {Vacuum},} {\protect\JournalTitle{Phys. Rev. Lett.}} \textbf{104}, 220404 (2010).

\bibitem{gonoskov_probing_2013}
A.~Gonoskov, I.~Gonoskov, C.~Harvey, \emph{et~al.}, \enquote{Probing {Nonperturbative} {QED} with {Optimally} {Focused} {Laser} {Pulses},} {\protect\JournalTitle{Phys. Rev. Lett.}} \textbf{111}, 060404 (2013).

\bibitem{robinson_interaction_2018}
A.~P.~L. Robinson and A.~V. Arefiev, \enquote{Interaction of an electron with coherent dipole radiation: {Role} of convergence and anti-dephasing,} {\protect\JournalTitle{Physics of Plasmas}} \textbf{25}, 053107 (2018).

\bibitem{gonoskov_charged_2022}
A.~Gonoskov, T.~Blackburn, M.~Marklund, and S.~Bulanov, \enquote{Charged particle motion and radiation in strong electromagnetic fields,} {\protect\JournalTitle{Rev. Mod. Phys.}} \textbf{94}, 045001 (2022).

\bibitem{stratton_diffraction_1939}
J.~A. Stratton and L.~J. Chu, \enquote{Diffraction {Theory} of {Electromagnetic} {Waves},} {\protect\JournalTitle{Phys. Rev.}} \textbf{56}, 99--107 (1939).

\bibitem{leshchenko_coherent_2015}
V.~E. Leshchenko, \enquote{Coherent combining efficiency in tiled and filled aperture approaches,} {\protect\JournalTitle{Opt. Express}} \textbf{23}, 15944 (2015).

\bibitem{vais_direct_2018}
O.~E. Vais and V.~Y. Bychenkov, \enquote{Direct electron acceleration for diagnostics of a laser pulse focused by an off-axis parabolic mirror,} {\protect\JournalTitle{Appl. Phys. B}} \textbf{124}, 211 (2018).

\bibitem{vais_characterizing_2020}
O.~E. Vais, A.~G.~R. Thomas, A.~M. Maksimchuk, \emph{et~al.}, \enquote{Characterizing extreme laser intensities by ponderomotive acceleration of protons from rarified gas,} {\protect\JournalTitle{New J. Phys.}} \textbf{22}, 023003 (2020).

\bibitem{khazanov_exawatt_2023}
E.~Khazanov, A.~Shaykin, I.~Kostyukov, \emph{et~al.}, \enquote{{eXawatt} {Center} for {Extreme} {Light} {Studies},} {\protect\JournalTitle{High Pow Laser Sci Eng}} \textbf{11}, e78 (2023).

\bibitem{christov_transmission_1985}
I.~P. Christov, \enquote{Transmission of radiation with rapid temporal modulation through optical instruments,} {\protect\JournalTitle{Opt Quant Electron}} \textbf{17}, 353--357 (1985).

\bibitem{zhao_investigation_2015}
Z.-x. Zhao, Y.-q. Gao, Y.~Cui, \emph{et~al.}, \enquote{Investigation of phase effects of coherent beam combining for large-aperture ultrashort ultrahigh intensity laser systems,} {\protect\JournalTitle{Appl. Opt.}} \textbf{54}, 9939 (2015).

\bibitem{born_principles_1999}
M.~Born, E.~Wolf, A.~B. Bhatia, \emph{et~al.}, \emph{Principles of {Optics}: {Electromagnetic} {Theory} of {Propagation}, {Interference} and {Diffraction} of {Light}} (Cambridge University Press, 1999), 7th ed.

\bibitem{soloviev_improving_2022}
A.~Soloviev, A.~Kotov, M.~Martyanov, \emph{et~al.}, \enquote{Improving focusability of post-compressed {PW} laser pulses using a deformable mirror,} {\protect\JournalTitle{Opt. Express}} \textbf{30}, 40584 (2022).

\bibitem{rondepierre_propagation_2023}
A.~Rondepierre, D.~Oumbarek~Espinos, A.~Zhidkov, and T.~Hosokai, \enquote{Propagation and focusing dependency of a laser beam with its aberration distribution: understanding of the halo induced disturbance,} {\protect\JournalTitle{Opt. Continuum}} \textbf{2}, 1351 (2023).

\bibitem{fu2024}
X.~Fu, Y.~Lei, H.~Li, \emph{et~al.}, \enquote{Optical alignment technology for 1-meter accurate infrared magnetic system telescope,} {\protect\JournalTitle{Journal of Astronomical Telescopes, Instruments, and Systems}} \textbf{10}, 014004--014004 (2024).

\end{thebibliography}

\begin{appendices}
\section{Calculating mirror parameters}
Let us consider the incidence of a laser beam on a parabolic mirror using the surface described by Eq.(\ref{Eq1}) as an example. The incident beam propagates in the negative direction of the \textit{z}-axis. The optical scheme corresponding to this configuration is shown in Fig.\ref{Fig1}. To model the mirror for a certain focusing scheme it is necessary to know the value of \textit{F} responsible for the focal length of the parabola (defined as $f=2F$/$(1+\psi_{off})$), which is usually given. However, at the beginning we only had some specific design requirements (described in Section \ref{sec:results1}) and beam width. We needed to recreate the configuration first and study it for various number of beams. Sometimes it is more convenient to proceed from other pre-determined or easily calculated parameters, such as beam incidence and convergence angles ($\alpha$ and $\varphi$ respectively), as well as the beam width \textit{d}. Therefore, one needs to additionally calculate  \textit{F}, based on these values. To do it, we can express the coordinates of the extreme points $P_1$ and $P_2$ for the beam on the mirror through the aforementioned parameters as

\begin{equation*}
x'_1=p, \quad z'_1=-\frac{p}{\tan\alpha}, \quad x'_2=p+d, \quad z'_2=-\frac{p}{\tan(\alpha+\varphi)}
\end{equation*}
Off-axis parameter \textit{p} is related to focal length \textit{F} due to the property of the parabola: all rays leaving the focus of the parabola and reflecting from it travel the same distance to the line perpendicular to the axis of the parabola and containing the focus (in Fig.\ref{Fig1}, such a line coincides with the \textit{x}-axis). Basing on this,

\begin{equation}
F(\alpha,p) = \frac{p}{2}\frac{\cos\alpha+1}{\sin\alpha}
\label{Eq2}
\end{equation}
Then we can substitute the coordinates of points $P_1$ and $P_2$ into Eq.(\ref{Eq1}), subtract the obtained expressions and use Eq.(\ref{Eq2}). That gives the following equation

\begin{equation}
AF^2+BFd+Cd^2=0
\label{Eq3}
\end{equation}
where the dimensionless factors \textit{A}, \textit{B} and \textit{C} are expressed through the angles $\alpha$ and $\varphi$ as follows:

\begin{equation*}
A=4 \left( \frac{2\cos\alpha-\cos(2\alpha)-1}{\sin^2\alpha}(\tan\alpha+\tan\varphi)+\frac{2(\cos\alpha-1)}{\sin\alpha}(1-\tan\alpha \tan\varphi) \right)
\end{equation*}
\begin{equation*}
B=\frac{\cos\alpha-1}{\sin\alpha}(\tan\alpha+\tan\varphi)-(1-\tan\alpha \tan\varphi)
\end{equation*}
\begin{equation*}
C=-(\tan\alpha+\tan\varphi)
\end{equation*}
The feasible solution of Eq.(\ref{Eq3}) is

\begin{equation}
F(\alpha,\varphi,d) = d\frac{-B+\sqrt{B^2-4AC}}{2A}
\label{Eq4}
\end{equation}

\section{Description of field simulation problem}
Let us formulate generally the numerical simulation problem devoted to reflection of an arbitrary laser beam from a mirror of arbitrary shape. We can define the shape of the surface in the form of a mapping

\begin{equation}
\vec{f}:\mathbb{R}^2 \rightarrow \mathbb{R}^3, \quad \vec{f}(x', y')\rightarrow\vec{r}=\{x,y,z\}
\label{Eq9}
\end{equation}
where $\vec{r}$ is a point of the reflecting surface in three-dimensional space, whereas $(x', y')$ is a point in the space of surface parameters. In the case of a parabolic surface, this mapping requires defining a set of unit vectors responsible for an auxiliary coordinate system on a mirror, and a set of parameters $\{d,\alpha,\varphi\}$ or $\{d,\alpha,p\}$. We can introduce tangent vectors to the surface at the point $(x', y')$

\begin{equation*}
\vec{\tau}_1(x', y')=\frac{\partial \vec{f}}{\partial x'}, \quad \vec{\tau}_2(x', y')=\frac{\partial \vec{f}}{\partial y'}
\end{equation*}
Then the normal to the surface at the same point is written as

\begin{equation*}
\vec{n}^*(x', y')=[\vec{\tau}_1 \times \vec{\tau}_2]=\left[\frac{\partial \vec{f}}{\partial x'} \times \frac{\partial \vec{f}}{\partial y'}\right]
\end{equation*}
which is not a unit vector, however, it allows to express the differential area dS through the product of the differentials of the surface parameters: $dS=|\vec{n}^*|dx'dy'$.

If the incident field is given at any point of space, meaning functions $\vec{E}_i(\vec{r})$ and $\vec{B}_i(\vec{r})$, responsible for the incident fields, are defined, then the reflected electric field at the point $\vec{r}_1$ can be calculated, using

\begin{dmath}
\vec{E}(\vec{r}_1) = \frac{1}{4\pi} \iint_S \left(2ik\left[\vec{n}^*(x', y') \times \vec{B}_i\left(\vec{f}(x', y')\right)\right] \psi\left(\vec{f}(x', y'), \vec{r}_1\right) + 2 \left(\vec{n}^*(x', y') \cdot \vec{E}_i\left(\vec{f}(x', y')\right)\right)\nabla\psi\left(\vec{f}(x', y'), \vec{r}_1\right) \right) \,dx' \,dy'
\label{Eq10}
\end{dmath}
which is based on Eq.(\ref{Eq7}). Here $S'$ is a preimage of surface \textit{S}  in a surface parameter space,  $\psi(\vec{r},\vec{r}_1)=\frac {e^{ik|\vec{r}-\vec{r}_1|}}{|\vec{r}-\vec{r}_1|}$  is the Green’s function and the operator $\nabla$ supposes the derivative to be taken with respect to  $\vec{r}$, meaning 

\begin{equation*}
\nabla\psi(\vec{r},\vec{r}_1)=(\vec{r}-\vec{r}_1) \frac {e^{ik|\vec{r}-\vec{r}_1|}}{|\vec{r}-\vec{r}_1|^2} \left( ik - \frac{1}{|\vec{r}-\vec{r}_1|}\right)
\end{equation*}
In a similar way, for magnetic field we get from Eq.(\ref{Eq8})

\begin{equation}
\vec{B}(\vec{r}_1) = \frac{1}{4\pi} \iint_S \left(2\left[\vec{n}^*(x', y') \times \vec{B}_i\left(\vec{f}(x', y')\right)\right] \times \nabla\psi\left(\vec{f}(x', y'), \vec{r}_1\right) \right) \,dx' \,dy'
\label{Eq11}
\end{equation}
The Eq.(\ref{Eq10}) and (\ref{Eq11}) allow to solve the numerical simulation problem devoted to reflection of an arbitrary laser beam from a mirror of arbitrary shape, analytically defined by Eq.(\ref{Eq9}).

\section{Influence of jitter}

\begin{figure}[!ht]
\centering
\includegraphics[width=10.3cm]{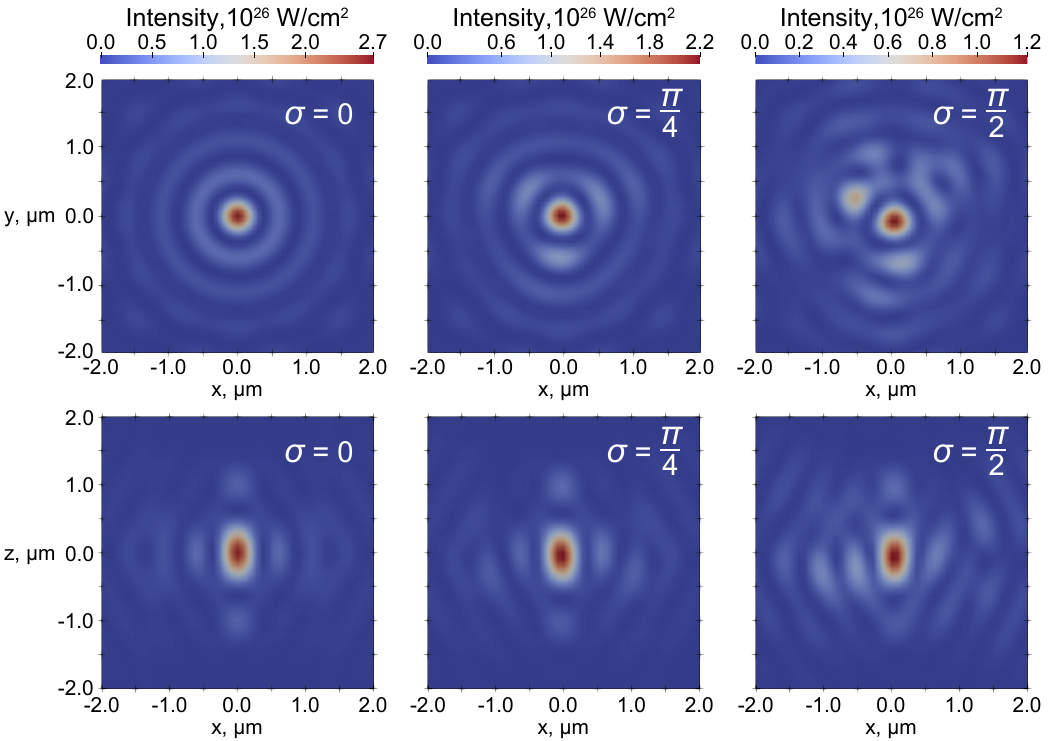}
\caption{Intensity distribution in focal area of a system with 12 beams for transversal (top) and longitudinal (bottom) planes for ideal phasing (left) and for different values of jitter, generated from a normal distribution with the corresponding value of $\sigma$ (middle and right), for the focusing scheme 1 (Fig.\ref{Fig2} (left))}
\label{fig:s1}
\end{figure}

\begin{figure}[ht!]
\centering
\includegraphics[width=10.3cm]{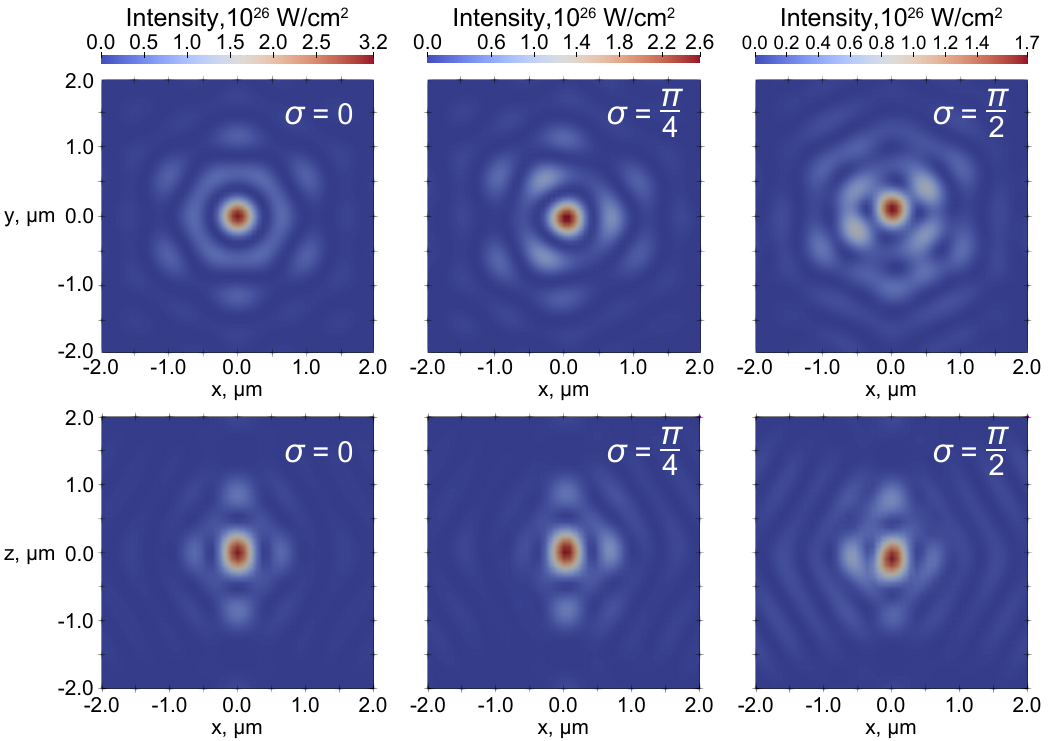}
\caption{Intensity distribution in focal area of a system with 12 beams for transversal (top) and longitudinal (bottom) planes for ideal phasing (left) and for different values of jitter, generated from a normal distribution with the corresponding value of $\sigma$ (middle and right), for the focusing scheme 2 (Fig.\ref{Fig2} (middle))}
\label{fig:s2}
\end{figure}

The presence of phase difference has a significant effect on the field and intensity distribution in the focal region (Fig.\ref{fig:s1} and \ref{fig:s2}). The phase difference not greater than $\frac{\pi}{6}$ in absolute value causes maximum intensity decrease by 10\%, for $\frac{\pi}{4}$ it is 20\%, and for $\frac{\pi}{2}$ it is close to 50\%.. Even without leading to a catastrophic drop in intensity, jitter value within $\frac{\pi}{4}$ does not noticeably change the shape of the spot, compared to the case of ideal phasing, while for jitter of $\frac{\pi}{2}$ there are noticeable distortions in the shape of the main spot and significant side maxima. These conclusions are the same for both focusing schemes studied.


\section{Influence of mirror alignment inaccuracies}

\begin{figure}[ht!]
\centering
\includegraphics[width=10.5cm]{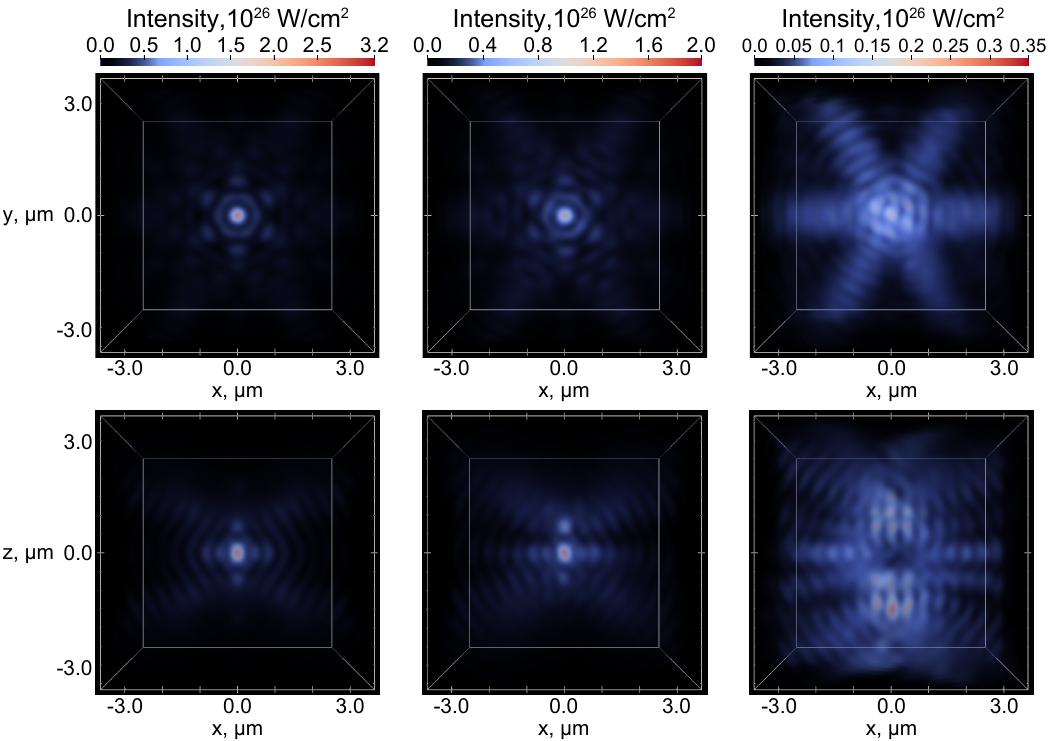}
\caption{Intensity distribution in focal area for transversal (top) and longitudinal (bottom) planes for reference configuration (left) and for various mirror shifts ($\sigma$=0,5 $\mu$m (middle) and $\sigma$=2 $\mu$m (right))}
\label{fig:s3}
\end{figure}

\begin{figure}[ht!]
\centering
\includegraphics[width=10.5cm]{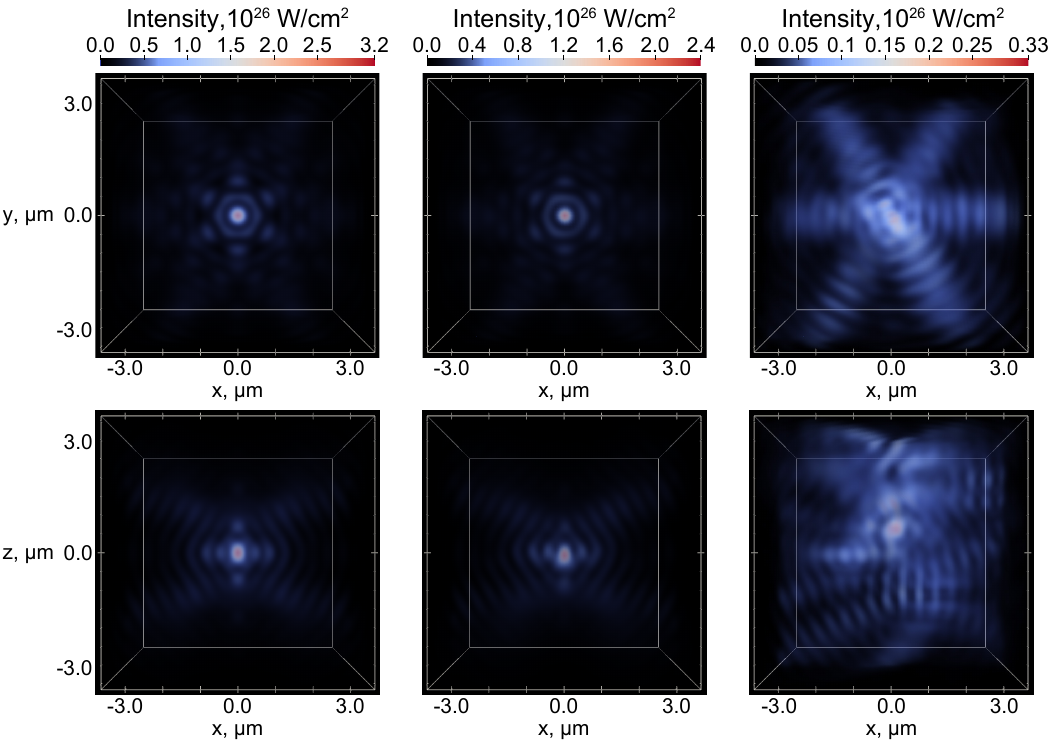}
\caption{Intensity distribution in focal area for transversal (top) and longitudinal (bottom) planes for reference configuration (left) and for various mirror rotations around \textit{y} axis ($\sigma$=5×$10^{-7}$ rad. (middle) and $\sigma$=5×$10^{-6}$ rad. (right))}
\label{fig:s4}
\end{figure}

The characteristic intensity distributions near the focus in the cases of shifts (vertical) and rotations (type 1) are also obtained (Fig.\ref{fig:s3} and \ref{fig:s4} accordingly). In these cases, the focal point moves because there has been a change in the position of the focusing element (parabolic mirror). Therefore, the intensity maximum at a large displacement can move far away from the original focal point of the reference case, which is (0,0,0).


\section{Influence of wavefront aberrations}

\begin{figure}[ht!]
\centering
\includegraphics[width=11.7cm]{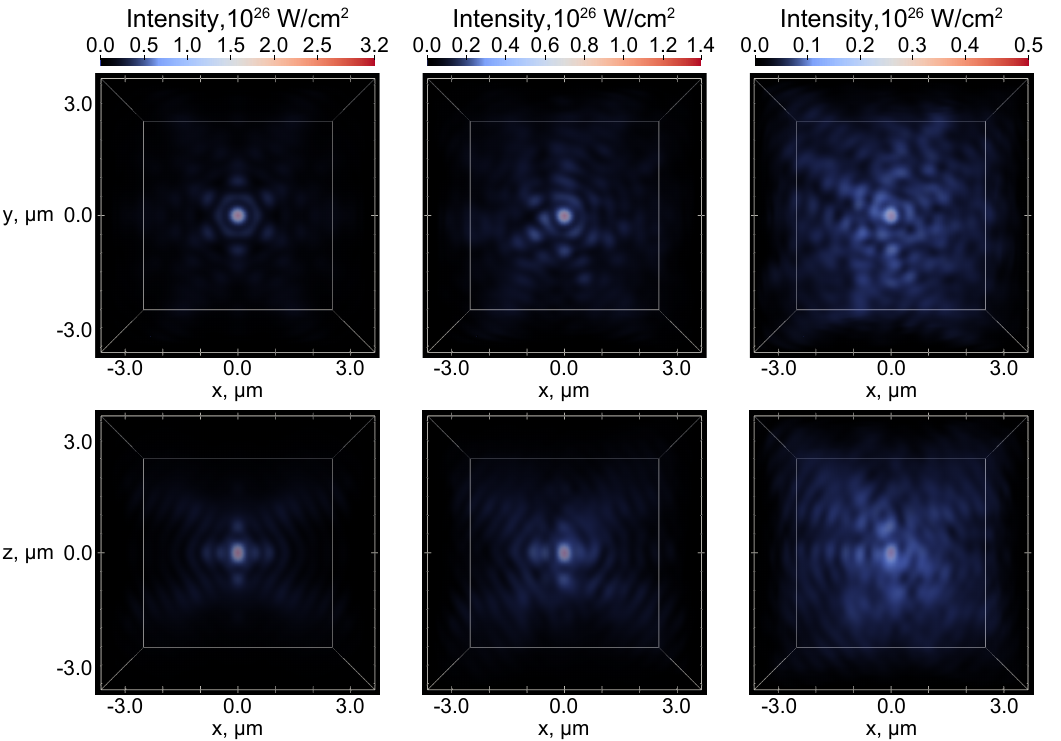}
\caption{Intensity distribution in focal area for transversal (top) and longitudinal (bottom) planes for reference configuration (left ), for $\sigma_{beam}\approx$ 0,83 rad. (middle) and $\sigma_{beam}\approx$ 1,41 rad. (right)}
\label{fig:s5}
\end{figure}

The characteristic intensity distributions in the focal area for wavefront aberrations of various strength are presented in Fig.\ref{fig:s5}. These results are related to the study of cumulative influence of the selected set of distortion harmonics. In this case, the mirrors are fixed and the aberrations do not change the beam that much in the central part (Fig.\ref{Fig8} (left)), so it is expected that the intensity maximum should not move far from the focal point of reference case, which is (0,0,0). The displacement of the intensity maximum is due to wavefront deformation, and the picture is more similar to the case of jitter (Fig.\ref{fig:s1} and \ref{fig:s2}), where the maximum predominantly remains at the center, than to the case of shifts and rotations (Fig.\ref{fig:s3} and \ref{fig:s4}).

The isolated effect of different type of aberrations was studied for a certain set of single Zernike polynomials $Z_n^m(\rho,\phi)$, every one of which corresponds to a certain aberration harmonic, listed in Table \ref{tab:functions}. To get the form of $Z_n^m(x_1,x_2 )$, that is used in the programming code, the coordinates are converted in a standard way $\rho^2=x_1^2+x_2^2$, $\rho \cos\phi=x_1$, $\rho \sin\phi=x_2$.

\begin{table}[htbp]
\centering
\caption{\bf Zernike polynomials for selected aberrations}
\begin{tabular}{ ccc>{\centering}m{10mm}>{\centering}m{10mm}c } 
\hline 
 \multicolumn{3}{c}{\multirow{2}{*}{\textbf{Aberration type}}} & \multicolumn{2}{c}{\textbf{Zernike indices}} & \multirow{2}{*}{\textbf{$Z_n^m(\rho,\varphi)$}} \\ \cline{4-5}
 \multicolumn{3}{c}{} &  \textbf{n} & \textbf{m} & \\ 
 \hline \hline
 \multicolumn{2}{c}{\multirow{2}{*}{Tilt}} & vertical & 1 & -1 &  $2\rho \sin\phi$\\ \cline{3-6}
 \multicolumn{2}{c}{} &  horizontal & 1 & 1 &  $2\rho \cos\phi$\\ \hline
 \multicolumn{3}{c}{Defocus} &  2 & 0 &  $\sqrt{3}(2\rho^2-1)$\\ \hline
 \multirow{2}{*}{Spherical} &  \multicolumn{2}{c}{primary} & 4 & 0 & $\sqrt{5}(6\rho^4-6\rho^2+1)$\\ \cline{2-6}
 & \multicolumn{2}{c}{secondary} & 6 & 0 & $\sqrt{7}(20\rho^6-30\rho^4+12\rho^2-1)$\\ \hline
 \multirow{4}{*}{Coma} &  \multirow{2}{*}{primary} & vertical & 3 & -1 &  $\sqrt{8}(3\rho^3-2\rho) \sin\phi$ \\ \cline{3-6}
 & & horizontal & 3 & 1 &  $\sqrt{8}(3\rho^3-2\rho) \cos\phi$ \\ \cline{2-6}
 & \multirow{2}{*}{secondary} & vertical & 5 & -1 & $\sqrt{12}(10\rho^5-12\rho^3+3\rho) \sin\phi$ \\ \cline{3-6}
 & & horizontal & 5 & 1 &  $\sqrt{12}(10\rho^5-12\rho^3+3\rho) \cos\phi$ \\ \hline
 \multirow{4}{*}{Astigmatism} &  \multirow{2}{*}{primary} & oblique & 2 & -2 &  $\sqrt{6}\rho^2 \sin(2\phi)$ \\ \cline{3-6}
 & & vertical & 2 & 2 &  $\sqrt{6}\rho^2 \cos(2\phi)$ \\ \cline{2-6}
 & \multirow{2}{*}{secondary} & oblique & 4 & -2 & $\sqrt{10}(4\rho^4-3\rho^2)\sin(2\phi)$ \\ \cline{3-6}
 & & vertical & 4 & 2 &  $\sqrt{10}(4\rho^4-3\rho^2)\cos(2\phi)$ \\ \hline
 \multicolumn{2}{c}{\multirow{2}{*}{Trefoil}} & vertical & 3 & -3 &  $\sqrt{8}\rho^3 \sin(3\phi)$\\ \cline{3-6}
 \multicolumn{2}{c}{} &  oblique & 3 & 3 &  $\sqrt{8}\rho^3 \cos(3\phi)$\\ \hline
\end{tabular}
  \label{tab:functions}
\end{table}

\begin{figure}[!ht]
\centering
\includegraphics[width=2.5cm]{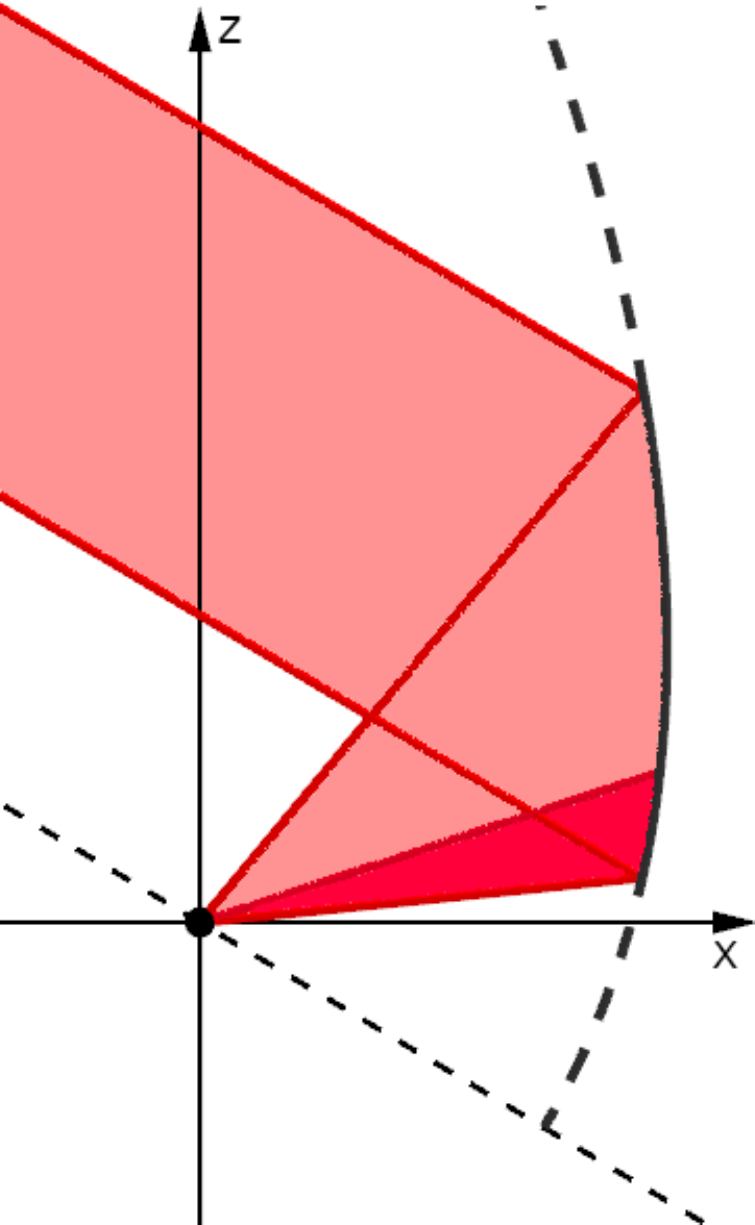}
\includegraphics[width=5cm]{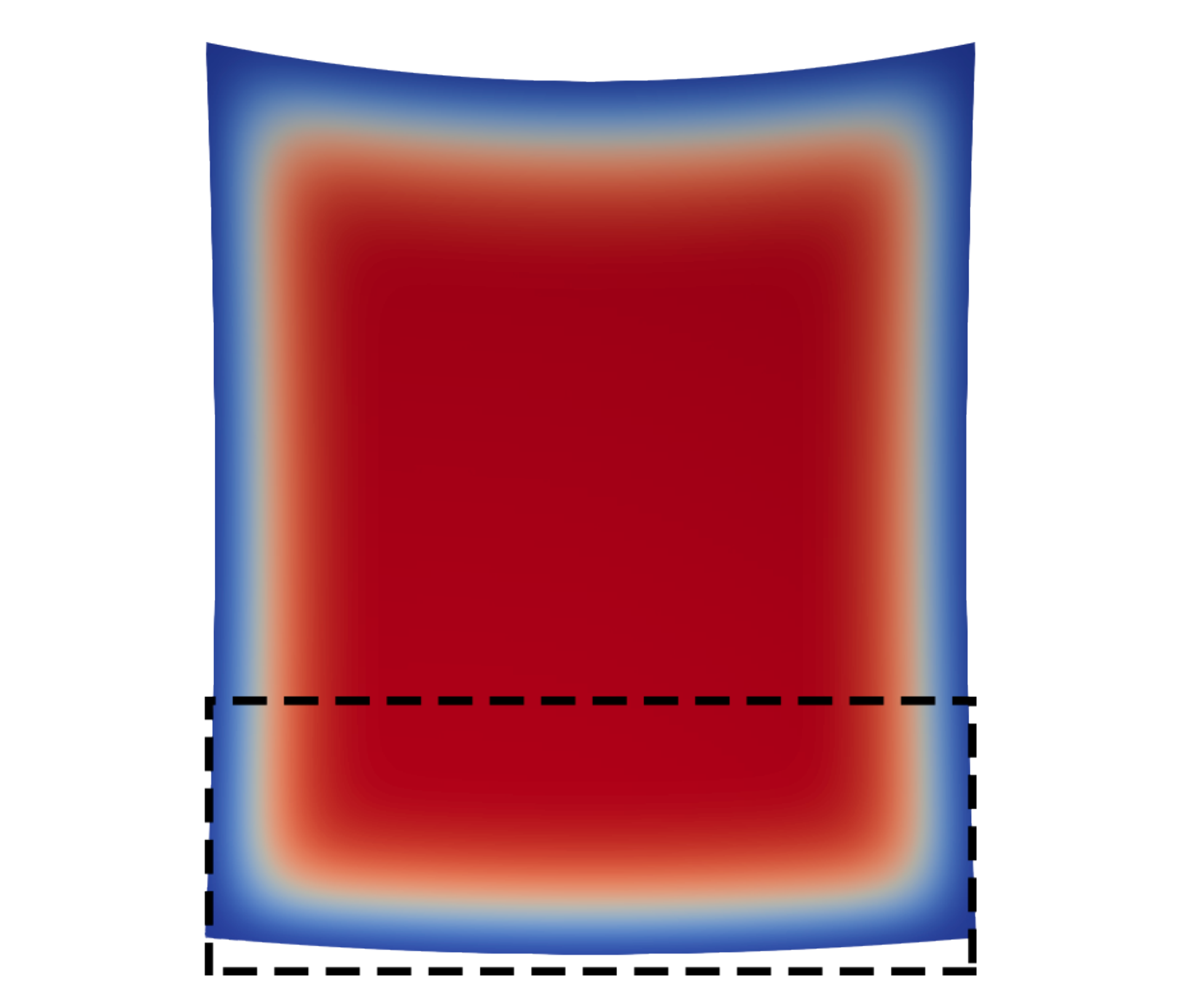}
\includegraphics[width=5.5cm]{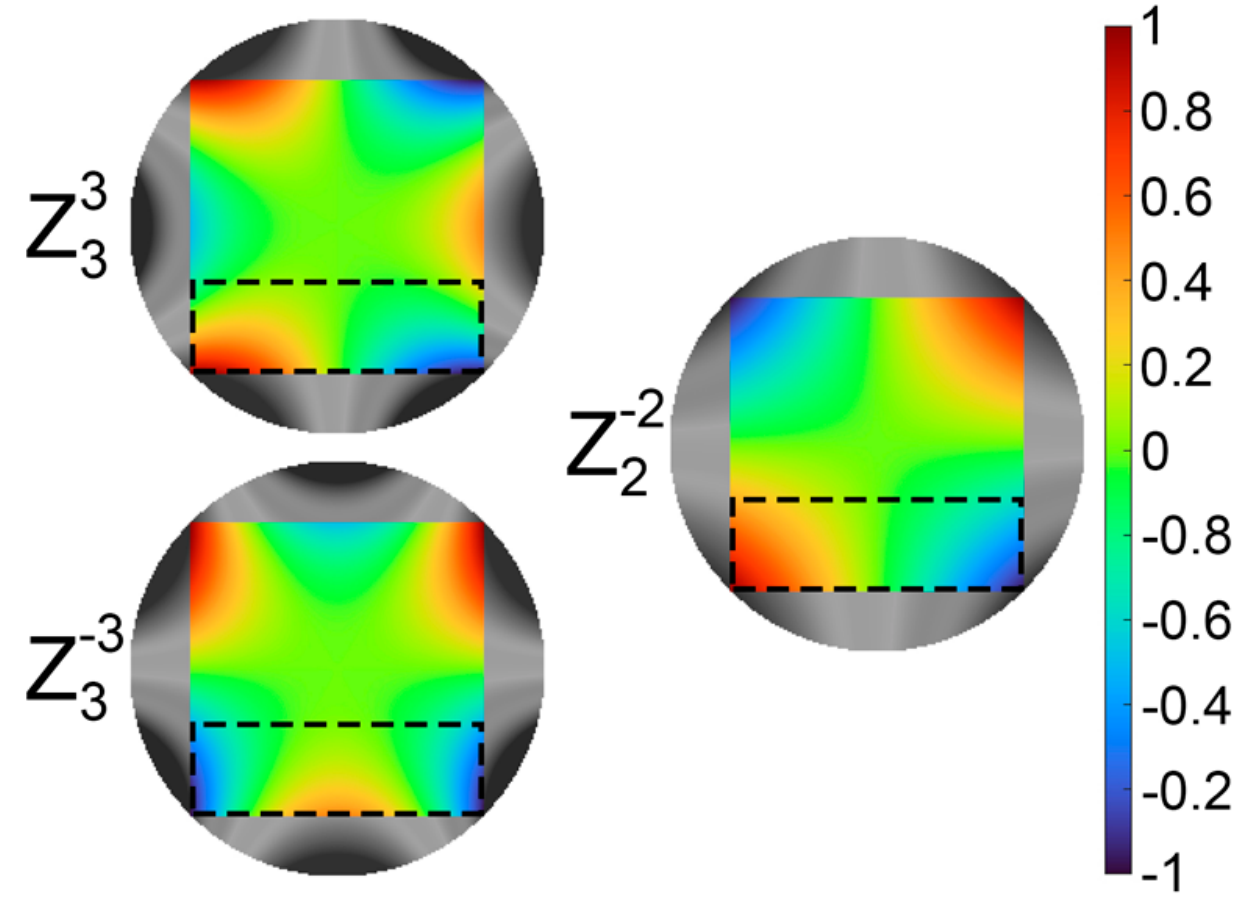}
\caption{Area of the beam, which contributes the most to the aggregate field (highlighted with dark red on the left and dashed line in the middle and on the right)}
\label{fig:11}
\end{figure}

Based on the results presented in Fig.\ref{Fig10}, the least influence on the coherent summation efficiency from the studied set is exerted by trefoil-type aberrations ($Z_3^{-3}$ and $Z_3^3$) and primary vertical astigmatism ($Z_2^{-2}$). Apparently, this is due to the main contribution to field at the focus of the system being given by the radiation reflected from a belt of small width at the lower boundary of the mirror (see Fig.\ref{fig:11}), since it has a polarization that differs from a vertical by the small angle, making it more optimal for coherent summation. The less significant contribution of the polynomials $Z_2^{-2}$, $Z_3^{-3}$ and $Z_3^3$ is attributed to the fact, that phase presumably remains unchanged even in the presence of aberrations for a more significant (compared to other polynomials) part of the described area of a beam.

\end{appendices}

\end{document}